\documentclass[runningheads]{llncs}
\usepackage{multirow}
\usepackage{graphicx}
\usepackage{hyperref}
\usepackage{url}
\usepackage{booktabs}
\usepackage{threeparttable}
\usepackage{color}

\usepackage[table,xcdraw]{xcolor}
\definecolor{secondary}{gray}{0.75}
\begin{document}
\title{Requirements Engineering using Generative AI: Prompts and Prompting Patterns}

%

\author{Krishna Ronanki\inst{1}\orcidID{0009-0001-8242-6771} \and
Beatriz Cabrero-Daniel\inst{1}\orcidID{0000-0001-5275-8372} \and
Jennifer Horkoff\inst{1}\orcidID{0000-0001-6794-9585} \and 
Christian Berger \inst{1}\orcidID{0000-0002-4828-1150}}

\authorrunning{Ronanki et al.}
%
\institute{University of Gothenburg, Gothenburg, Sweden}

\maketitle             
\begin{abstract}

\textbf{[Context]} Companies are increasingly recognizing the importance of automating Requirements Engineering (RE) tasks due to their resource-intensive nature. The advent of GenAI has made these tasks more amenable to automation, thanks to its ability to understand and interpret context effectively.
\textbf{[Problem]} However, in the context of GenAI, prompt engineering is a critical factor for success. Despite this, we currently lack tools and methods to systematically assess and determine the most effective prompt patterns to employ for a particular RE task.
\textbf{[Method]} Two tasks related to requirements, specifically requirement classification and tracing, were automated using the GPT-3.5 turbo API. The performance evaluation involved assessing various prompts created using 5 prompt patterns and implemented programmatically to perform the selected RE tasks, focusing on metrics such as precision, recall, accuracy, and F-Score.
\textbf{[Results]} This paper evaluates the effectiveness of the 5 prompt patterns' ability to make GPT-3.5 turbo perform the selected RE tasks and offers recommendations on which prompt pattern to use for a specific RE task. Additionally, it also provides an evaluation framework as a reference for researchers and practitioners who want to evaluate different prompt patterns for different RE tasks.

\end{abstract}

\section{Introduction} \label{sec:int}

Researchers in Requirements Engineering (RE) have been exploring the use of Machine Learning (ML) and Deep Learning (DL) methods for various RE tasks, including requirements classification, prioritization, tracing, ambiguity detection, and modelling~\cite{alhoshan2023zero}. However, the majority of existing ML/DL approaches are based on supervised learning, which requires huge amounts of task-specific labelled training data. But, the lack of open-source RE-specific labelled data makes it difficult for RE researchers and practitioners to develop, train and test advanced ML/DL models~\cite{alhoshan2023zero} for their effective usage in RE tasks. 

Utilizing pre-trained Generative Artificial Intelligence (GenAI) models like Large Language Models (LLM) for performing RE tasks removes the need for large amounts of labelled data. Pre-trained LLMs are also observed to increase developer productivity~\cite{peng2023impact} and reduce code complexity~\cite{10.1145/3524842.3528470} among other things. The adoption of these LLMs in practical settings is also growing thanks to the development of IDE-integrated services around them~\cite{white2023chatgpt}. Recent LLMs are demonstrating increasingly impressive capabilities when performing a wide range of tasks~\cite{brown2020language}. These capabilities can be further improved in NLP tasks just by carefully crafting the input given to the model~\cite{haque2022think}. 

To interact with an LLM, one typically provides instructions written in Natural Language (NL)~\cite{white2023chatgpt}, referred to as prompts. The emerging practice of utilizing carefully selected and composed NL instructions to achieve the desired output from a GenAI model such as a pre-trained LLM is called prompt engineering~\cite{10.1145/3491102.3501825}. However, the problem with NL instructions is that they can be ambiguous in some contexts, i.e. some of the words in the instruction can have multiple interpretations which vary according to the context. One of the major risks of having an instruction that can be interpreted in different ways is that the LLM might interpret the instruction in a way that is different from the users'. This might lead to an output generation that is  considered unexpected and/or undesirable by the user. Despite the existence of empirically validated prompt engineering techniques like zero-shot, few-shot~\cite{brown2020language}, and chain-of-thought prompting~\cite{NEURIPS2022_9d560961}, prompt engineering is still more of an art than science. Even the order in which the samples are provided in a few-shot setting can make the difference between near state-of-the-art and random guess performance~\cite{lu2021fantastically}. 

One potential way to mitigate the challenges with the ambiguous nature of current prompt engineering practices is to develop a structured approach for crafting the NL prompts. However, to define and establish a prompt structure that can be generalised, it is crucial to identify any underlying patterns within the prompts that consistently generate desirable output for a given task. Different prompts need to be tested in order to see what different outputs they may lead to, thereby discovering any discrepancies between what is assumed or expected and what is understood by the models~\cite{cheng2023prompt}. This led to the development of prompt patterns. Prompt patterns can be defined as codified reusable patterns that can be applied to the input prompts to improve the desirability of the generated output while reducing the gap between the user's expected output and the model's generated output~\cite{white2023prompt}. 

Recognizing prompt patterns that consistently produce desirable output so that practitioners can leverage the numerous benefits of using these LLMs in RE tasks has significant value. Our work investigates and presents which prompt patterns can be used to produce outputs that are desirable and conform to users' expectations when using LLMs for two RE tasks. We chose the GPT 3.5-turbo model (which we will refer to as 'the model' from now on for the rest of the paper) as our choice of LLM which we accessed through the API to perform the RE tasks. The two RE tasks we chose to focus on in this study are a) binary requirements classification and b) identifying requirements that are dependent on each other (requirements traceability task). We measure the performance of the patterns' implementation in these tasks using measures like precision, recall, accuracy and F-score. We evaluate the performance of the model at different temperature settings to understand its effect on prompt patterns' performance. We recommend prompt patterns that achieve the best performance scores for the selected RE tasks in our experimental configuration. We also propose a framework to evaluate a prompt pattern's effectiveness for any RE task based on the methodology we employed for this study. To that end, we aim to answer the following research questions:

\vspace{0.2cm}
\noindent\fbox{%
\begin{minipage}{.96\columnwidth}

\textbf{RQ1}- What prompt patterns can be recommended for RE researchers and practitioners for binary requirements classification and requirement traceability tasks?

\textbf{RQ2}- How to evaluate a prompt pattern's effectiveness in performing any RE task?
\end{minipage}
}\\

\section{Background}

Our approach towards designing the binary requirements classification and requirements traceability tasks for our experimental setup involves influences from requirements Information Retrieval (IR). In the context of requirements IR, classification involves categorising requirements into different groups~\cite{zhang2023empirical}. Requirements classification and IR share common principles related to information organisation, search \& retrieval, and semantic understanding among other things. IR methods are also used to search for specific traceability information during software development, helping stakeholders locate related artefacts and trace the relationships between them.

\subsection{Prompt Engineering}

Prompt engineering~\cite{10.1145/3491102.3501825} or prompt programming~\cite{10.1145/3411763.3451760} or prompting is an emerging practice in which carefully selected and composed sentences are used to achieve the desired output (the process of engineering a natural language prompt). Prompt engineering allows the model's users to express their intent in plain language, rather than a specially designed programming language~\cite{10.1145/3544549.3585737}. Prompt engineering is still more of an art than science but there are few techniques that have been empirically validated in experimental settings to improve the GenAI model's performance.

\subsubsection{Zero-shot prompting}

In its most basic form, zero-shot prompting involves using natural language sentences to convey the ``problem to be solved'' or the ``expected output'', without providing any examples~\cite{10.1145/3544549.3585737}. It is a technique modelled after Zero-Shot Learning (ZSL), which directly applies previously trained models to predicting both seen and unseen classes without using any labelled training instances~\cite{lampert2009learning}. 

\subsubsection{Few-shot prompting}

Few-shot prompting builds on zero-shot prompting by conveying the ``problem to be solved'' or the ``expected output'' using a few demonstrations of the task (examples) at inference time as conditioning~\cite{radford2019language}.

\subsubsection{Chain-of-thought prompting}

Chain-of-thought prompting is a technique that is observed to enhance the reasoning capabilities of an LLM. Using this technique, the user constructs the prompts in a way that makes the model generate a coherent series of intermediate reasoning steps that lead to the final answer for the task at hand~\cite{NEURIPS2022_9d560961}. 

\subsection{Generative AI Model Temperature}

Temperature of a GenAI model is a parameter that controls the randomness of the model's output. When adjusting the temperature setting of a GenAI model, which ranges from 0.0 to 1.0, one is essentially controlling the randomness of the model's responses. Higher temperatures (e.g., 0.8 or 1.0) result in more diverse and ``creative'' responses, while lower temperatures (e.g., 0.2 or 0.5) produce more focused and deterministic responses. In our study, we evaluate the performance of the model in performing the selected RE tasks while implementing all 5 patterns over 3 temperature settings. The 3 temperature settings are 0.0(lowest), 0.4 (default) and 1.0(highest).

\subsection{Related Work} \label{sec:rw}

There are several works in the existing literature that focus on the application of LLMs for various IR-specific software engineering (SE), while few focus on RE tasks like binary and multi-class classification of requirements. Zhang et al.~\cite{zhang2023empirical} empirically evaluate ChatGPT’s performance on requirements IR tasks. Under the zero-shot setting, their results reveal that ChatGPT’s performance in IR tasks has high recall but low precision. They posit their evaluation provides preliminary evidence for designing and developing more effective requirements IR methods based on LLMs. Alhoshan et al.~\cite{alhoshan2023zero} report an extensive study using the contextual word embedding-based zero-shot learning (ZSL) approach for requirements classification. The study tested this approach by conducting more than 360 experiments using four language models with a total of 1020 requirements and found generic language models trained on general-purpose data perform better than domain-specific language models under the zero-shot learning approach. Their results show that ZSL achieves F-Scores from 66\% to 80\% for binary and multi-class classification tasks.

To the best of the authors' knowledge, there are no works that focus on measuring the performance of LLMs in requirements traceability tasks. A Systematic Mapping Study (SMS) performed by Li et al.~\cite{liapplications} presents 32 Machine Learning (ML) technologies and seven enhancement strategies for establishing trace links through their work. Their results indicate that ML technologies show promise in predicting emerging trace links by leveraging existing traceability information within the requirements. They identified three studies that they classified under the ``semantically similar words extraction'' enhancement strategy.

White et al.~\cite{white2023prompt} present prompt design techniques for software engineering in the form of patterns to automate common software engineering activities using LLMs. Prompt patterns serve as a means of knowledge transfer, similar to software patterns, by offering reusable solutions to common problems related to generating output and engaging with LLMs. They establish a framework for documenting prompt patterns that can be adapted to various domains, providing a systematic approach for structuring prompts to tackle a range of issues. The academic literature presents and discusses a catalogue of patterns that have proven successful in enhancing the quality of LLM-generated conversations.

Despite these studies, to our knowledge, there are no works that focus on measuring the performance of an LLM in performing a certain RE task while using a specific prompt pattern to craft your input prompts. This is crucial since there is an opportunity to identify prompt patterns that work better in comparison to others for particular RE tasks.

\section{Methodology} \label{sec:method}

\subsection{Experiment Design} \label{subsec:design}

We selected five prompt patterns for our experiments with the RE tasks: 1. Cognitive Verifier; 2. Context Manager; 3. Persona; 4. Question Refinement and 5. Template, out of 16 patterns presented by White et al.~\cite{white2023prompt}. They have been selected on the basis of the descriptions for each prompt pattern provided by the authors, which include the intent and motivation behind the pattern's proposal, the structure and key ideas that the pattern represents, an example implementation of the pattern in practice, and the observed consequences of the pattern's implementation in practice. We used these five patterns to craft prompts for the selected RE tasks and presented these prompts in Table~\ref{tab:table-1}. We performed each experiment five times using each of the prompts presented in Table~\ref{tab:table-1}, gathering the model's replies and computing the aggregated performance measures (precision, recall, accuracy and F-score) for each run.

\begin{table}[ht!]
\centering
\begin{tabular}{|p{1.6cm}|p{5.2cm}|p{5.2cm}|}
\hline
\textbf{Pattern} & \textbf{Classification} & \textbf{Tracing} \\ \hline \hline
\textbf{Cognitive Verifier} & Classify the given list of requirements into functional (labelled as F) and non-functional requirements (labelled as NF). Ask me questions if needed to break the given task into smaller subtasks. All the outputs to the smaller subtasks must be combined before you generate the final output. & List the IDs of requirements that are related to the [deprecated] feature in the requirements specification document below. Ask me questions if needed to break down the given task into smaller subtasks. All the outputs to the smaller subtasks must be combined before you generate the final output. \\ \hline
\textbf{Context Manager} & Classify the given list of requirements into functional (labelled as F) and non-functional requirements (labelled as NF). When you provide an answer, please explain the reasoning and assumptions behind your response. If possible, address any potential ambiguities or limitations in your answer, in order to provide a more complete and accurate response. & List the IDs of requirements that are related to the [deprecated] feature from the requirements specification document below. When you provide an answer, please explain the reasoning and assumptions behind your response. If possible, address any potential ambiguities or limitations in your answer in order to provide a more complete and accurate response. \\ \hline
\textbf{Persona} & Act as a requirements engineering domain expert and classify the given list of requirements into functional (labelled as F) and non-functional requirements (labelled as NF). & Act as a requirements engineering domain expert and list the IDs of requirements that are dependent on the [deprecated] feature in the following requirements specification document: \\ \hline
\textbf{Question Refinement} & Classify the given list of requirements into functional (labelled as F) and non-functional requirements (labelled as NF). If needed, suggest a better version of the question to use that incorporates information specific to this task and ask me if I would like to use your question instead. & List the IDs of requirements that are related to the [deprecated] feature from the requirements specification document below. If needed, suggest a better version of the question to use that incorporates information specific to this task and ask me if I would like to use your question instead. \\ \hline
\textbf{Template} & Read the following list of requirements and return the IDs of non-functional requirements only. Write the result as a list like: (ID=X) (ID=Y) (ID=Z) where X, Y, and Z are IDs of non-functional requirements. & List the IDs of requirements that are related to the [deprecated] feature in the requirements specification document below. Follow the provided template when generating the output: ID list: X.X.X.X; X.X.X; X.X.X.X etc. \\
\hline
\end{tabular}
\caption{Prompts Table}
\label{tab:table-1}
\end{table}

\subsection{Datasets} \label{subsec:datasets}

We used the PROMISE dataset~\cite{jane_cleland_huang_2007_268542} for the classification task and the PURE dataset for the requirements traceability task as they have been widely used in literature. The PROMISE dataset, which is available in .arff file format was converted into .csv format. The CSV file (ground truth for our experiment) had a total of 621 requirements, out of which 253 requirements were functional requirements labelled as F and 368 were non-functional requirements labelled as NF.

The PURE dataset~\cite{8049173} is composed of public requirements documents retrieved from the Web. The documents cover multiple domains, have different degrees of abstraction, and range from product standards to documents of public companies, to university projects. A general XML schema file (XSD) was also defined to represent these different documents in a uniform format. From this dataset, we chose a subset of the dataset where each requirement in the System Requirements Specification (SRS) documents has a numerical ID in the format of X.X.X.X. These IDs (trace links as we will call them from now on) were used to establish a reference point, helping identify which requirements in the SRS documents are connected or dependent on one another. For example, if requirement A contains any reference to requirement B, then requirement A also had a trace link referencing requirement B in the X.X.X.X format. We used this information to construct the ground truth for the traceability task. 

\subsection{Tasks} \label{subsec:tasks}
We begin the study by conducting a series of controlled experiments with the model for two tasks: (i) Binary Classification of Functional Requirements (FR) \& Non-Functional Requirements (NFR), and, (ii) Requirements Traceability, described in more detail below.

\subsubsection{Binary Classification of Functional and Non-Functional Requirements}

The task aims to distinguish Functional Requirements (FR) from Non-Functional Requirements(NFR), assuming that a requirement belongs to either the FR or NFR class. The PROMISE NFR dataset~\cite{jane_cleland_huang_2007_268542} was used for this purpose. The process we followed is:

\begin{enumerate}
    \item We wrote a Python program that randomly picks 50 requirements from a CSV file. This CSV file contains 621 unlabelled requirements, the same 621 requirements that make up the PROMISE dataset.
    \item We also input a prompt that we created using one of five patterns right into the program.
    \item We use the model through the API to perform the classification task. We do this five times and each time, the program chooses a different set of 50 requirements randomly.
    \item The program then automatically compares the classification results with the ground truth results from the PROMISE dataset.
    \item We repeated this whole process five times, once for each of the prompt patterns we were testing.
\end{enumerate}

\subsubsection{Requirements Traceability}

For this task, we had two sets of software requirement specifications (SRS) taken from the PURE dataset~\cite{8049173}: one for a home temperature control system called ``THEMAS'', and the other for defining a game interface and its functionalities, known as ``QHEADACHE''. The process we followed is:

\begin{enumerate}
    \item For consistency, we manually formatted these SRS files and removed any unnecessary information, like hyperlinks.
    \item We then created modified versions of these documents, referred to as ``THEMAS clean'' and ``QHEADACHE clean'', where we removed trace links.
    \item We provided these cleaned documents, without trace links, to the model programmatically.
    \item The model was given a requirement (randomly selected from the input documents) and asked to identify all related or dependent requirements.
    \item We repeated this process five times for each prompt pattern, with a different randomly selected set of requirements each time.
    \item This entire procedure was repeated five times, once for each prompt pattern we were testing.
\end{enumerate}

\subsection{Performance Metrics} \label{subsec:metrics}

The RE field has often adopted IR's measures, i.e., precision, recall, and the F-measure, to assess the effectiveness of any tool~\cite{8049186}. Since both tasks selected and defined for our study are related to requirements IR, we also used precision, recall, F-Score and accuracy to measure the performance of the model in performing both the RE tasks using the five prompt patterns. These measures were computed programmatically by comparing the model's outputs with the ground truth.

Consider a scenario where an analyst wants to identify all NFRs in a specification. In this scenario, a high recall indicates that the majority of the NFRs selected were accurately categorized as NFRs. Conversely, a low recall suggests that the majority of requirements were misclassified, with FRs being mistakenly identified as NFRs. A high precision signifies that most of the requirements classified as NFRs by the LLM are indeed NFRs. On the contrary, a lower precision suggests that a number of requirements identified as NFRs by the LLM are, in fact, FRs.

Consider a scenario where a requirement is marked as “deprecated”. In this scenario, it is important to trace all affected dependencies, i.e. find all requirements associated with the deprecated requirement. In this context, a high recall would signify that the majority of the associated or dependent requirements have been appropriately identified. Conversely, a low recall would indicate that only a limited number of the connections have been recognized. Precision provides a measure of the accuracy of retrieved links in relation to the dataset. A high precision indicates that a significant portion of the retrieved results indeed align with the dataset’s true links. On the other hand, a low precision implies that a considerable number of the retrieved results are not truly linked and should not have been flagged.

The F-score is the harmonic mean of precision and recall that takes both false positives and false negatives into account. It combines both of them into a single metric to provide a balanced evaluation of a model's performance.

\subsection{Threats to Validity}

While this study provides an analysis of prompt patterns in the context of binary requirements classification and tracing dependent requirements using LLMs, it is essential to acknowledge and address potential threats to the validity of the findings:

\textbf{Internal Validity:} There is a possibility for the existence of a degree of uncertainty in the ground truth since the data in our chosen datasets are labelled by humans(contributing authors of the dataset). Inter-rater variability and potential labelling bias could impact the reliability of performance metrics. The preprocessing steps applied to the data, such as cleaning, could influence the model's ability to capture complex patterns as well. The prompts crafted and presented in Table~\ref{tab:table-1} using the prompt patterns are also subject to the authors' capabilities and competence with the task at hand.

\textbf{External Validity:} Given the unique nature of some RE tasks, the findings from this study may not always generalize beyond binary requirements classification and tracing dependant requirements. 

\textbf{Construct Validity:} The findings of this study may be constrained by the characteristics and representativeness of the datasets in use. Since we used a GPT model, there is a possibility that the datasets we used could have been part of the training data for the GPT model. Next, the choice of prompt patterns is a critical aspect of this study. The selected patterns may not fully encompass the spectrum of possible patterns, potentially leading to an incomplete representation of LLM performance for the selected RE tasks.

\section{Results \& Analysis} \label{sec:results}

This section presents the aggregated results from the described experiments. Subsection~\ref{subsec:rq1} presents the performance measures of the model in performing the two RE tasks using the five selected prompt patterns. Subsection~\ref{subsec:rq2} presents recommendations for RE researchers and practitioners on which prompt patterns to use for the selected RE tasks based on our analysis of the results obtained, answering \textbf{RQ1}. We abstract our methodology and present it as a framework to evaluate the performance of any prompt pattern for a chosen RE task in Subsection \ref{subsec:rq3}, answering \textbf{RQ2}. 

\subsection{Prompt Patterns' Performance for the Selected RE Tasks} \label{subsec:rq1}

\renewcommand{\arraystretch}{1.3}
\begin{table*}[htbp]
	\centering
	\begin{threeparttable}
 \resizebox{\textwidth}{!}{
		\begin{tabular}{p{0.3\textwidth}p{0.15\textwidth}p{0.15\textwidth}p{0.15\textwidth}p{0.15\textwidth}}
			\toprule 
			\textbf{Prompt Pattern} & \textbf{Precision} & \textbf{Recall}  & \textbf{F-Score} & \textbf{Accuracy} \\
			\hline 
		Cognitive Verifier & \begin{minipage}{.9\textwidth} \includegraphics[width=15mm]{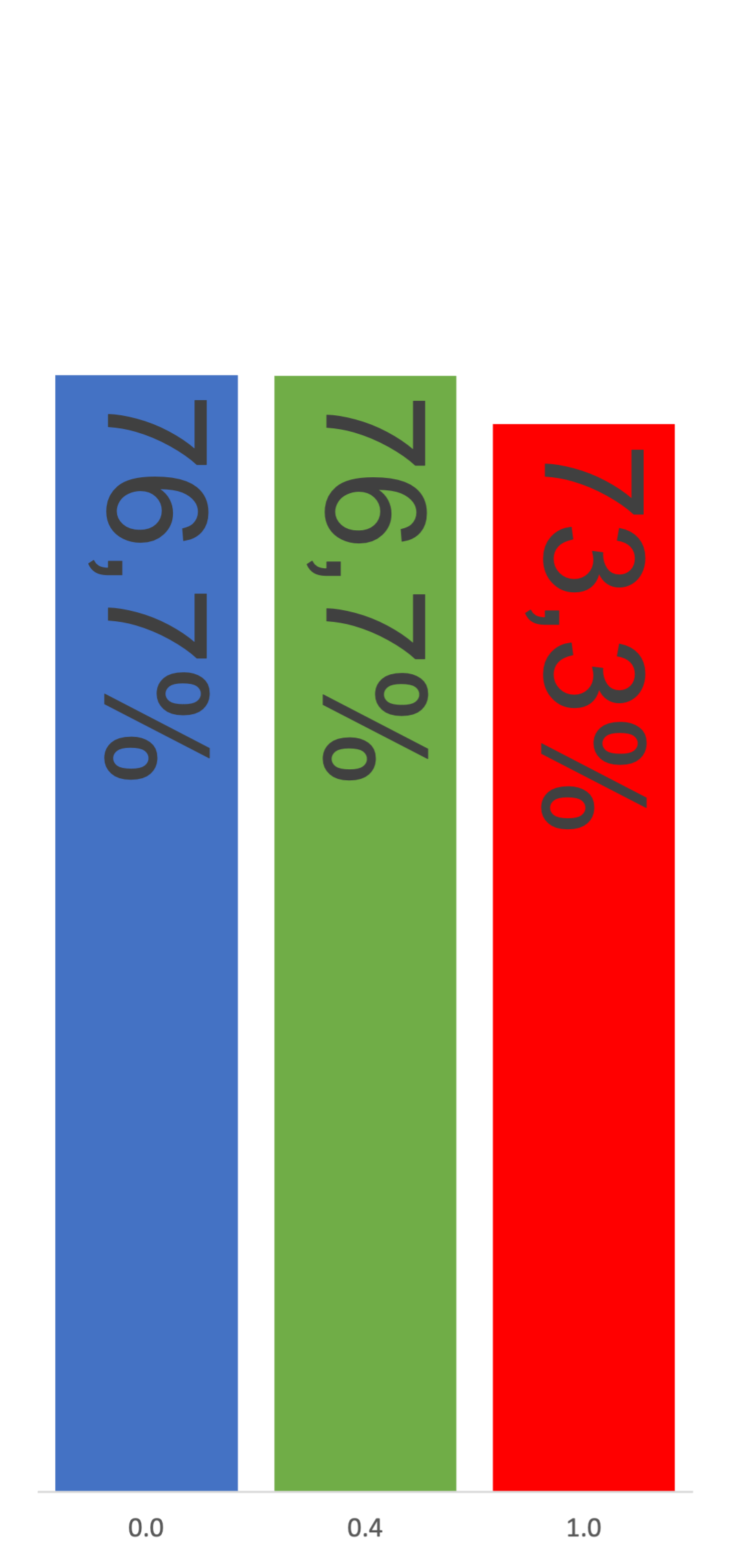} \end{minipage} 
			& \begin{minipage}{.2\textwidth} \includegraphics[width=15mm]{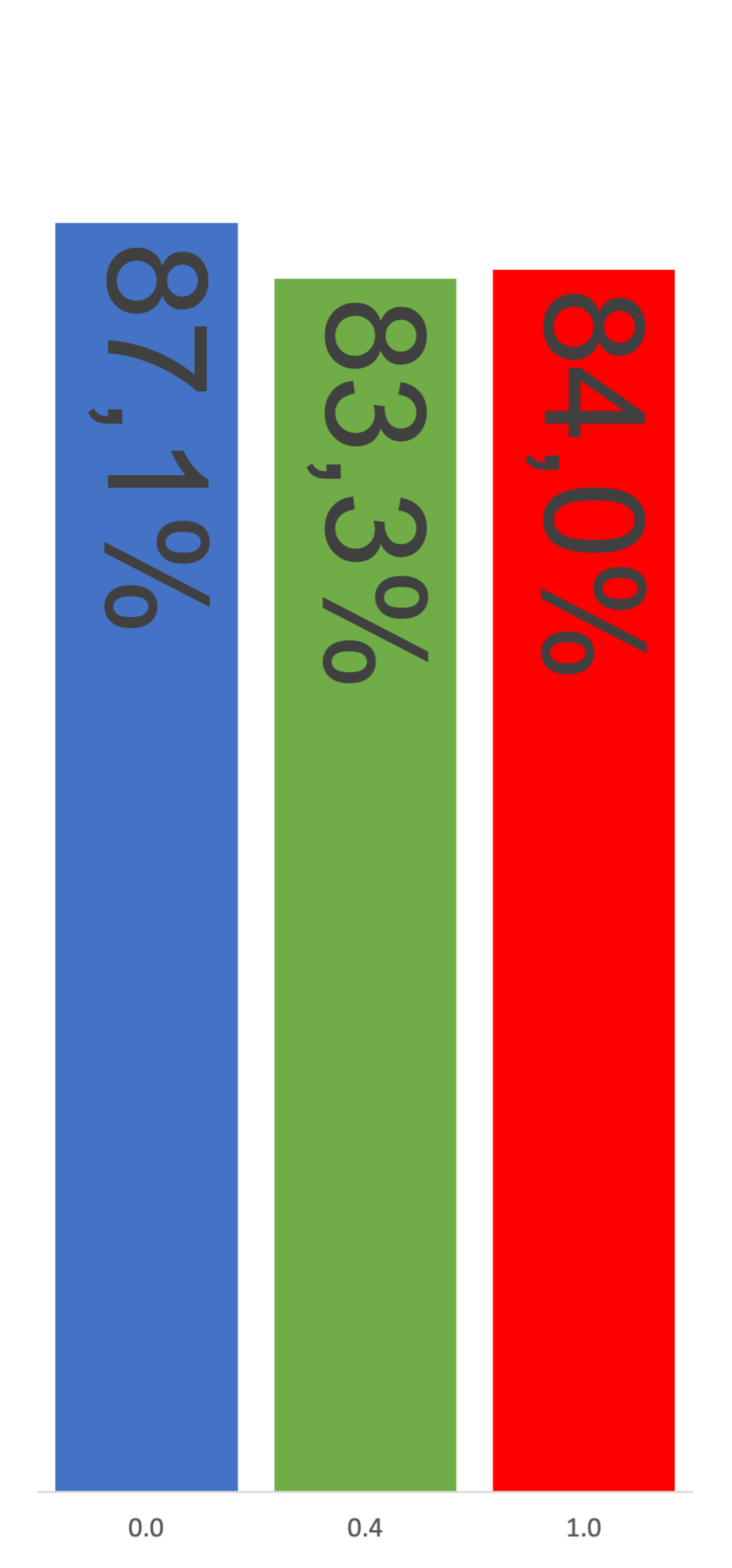} \end{minipage}
			& \begin{minipage}{.2\textwidth} \includegraphics[width=15mm]{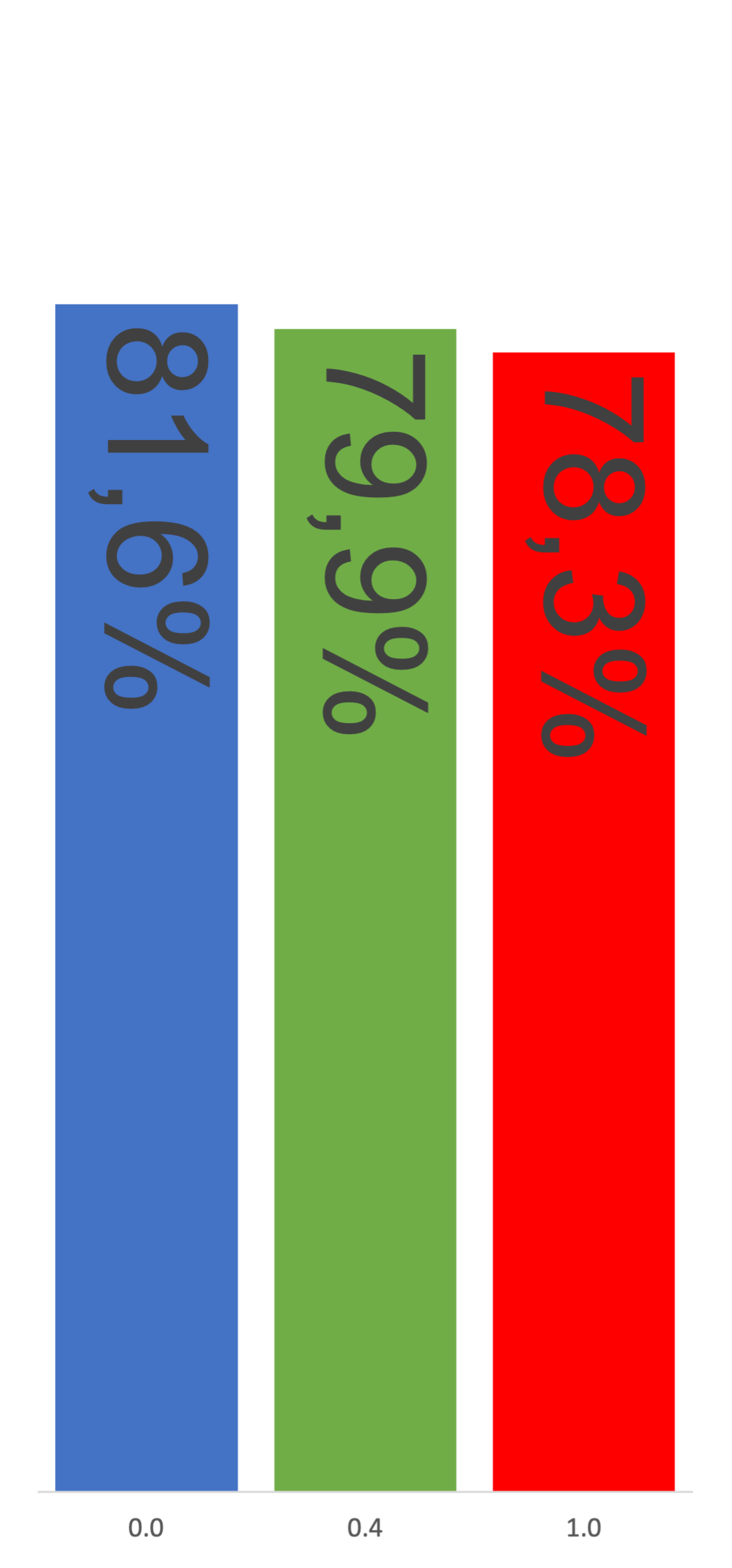} \end{minipage}
			& \begin{minipage}{.2\textwidth} \includegraphics[width=15mm]{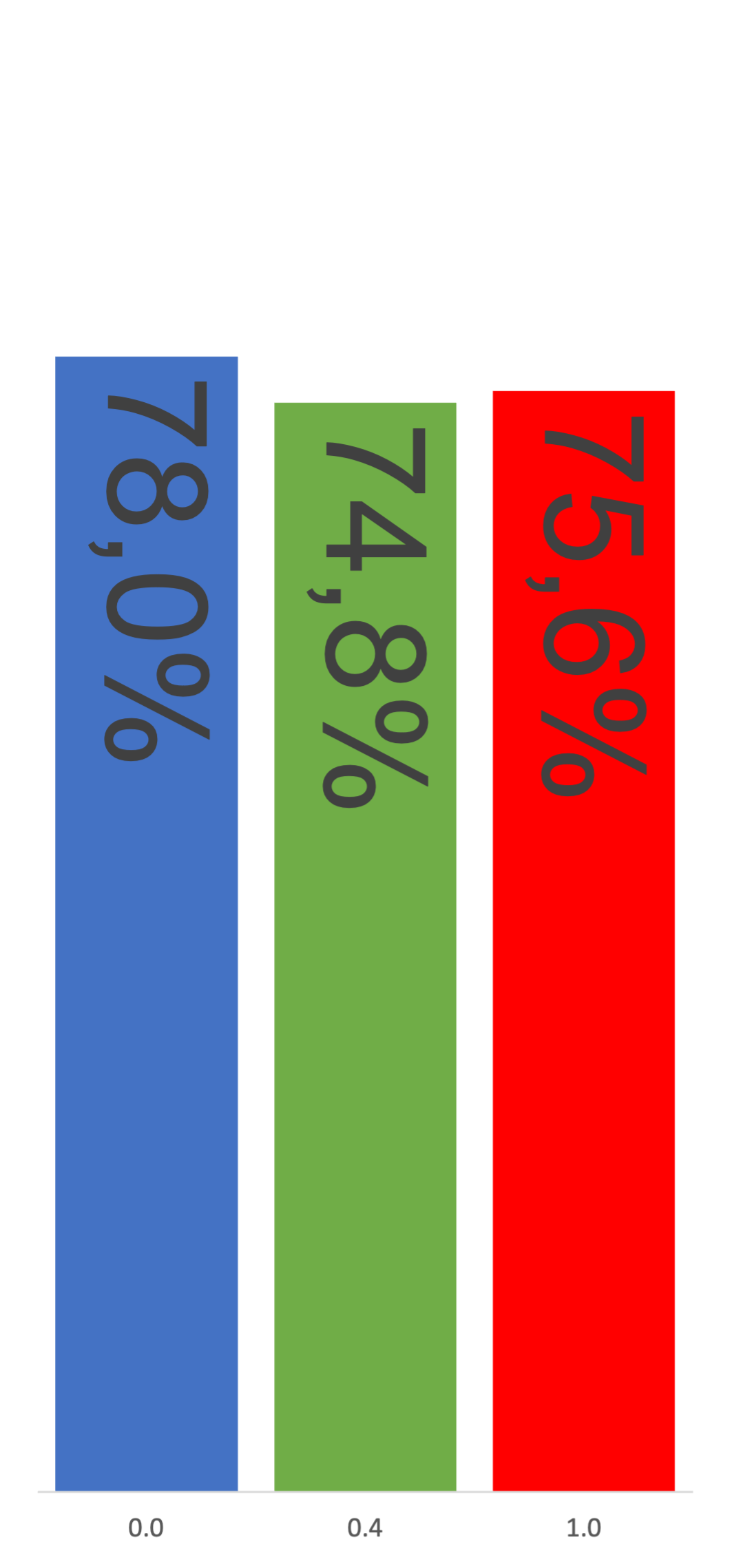} \end{minipage}\\
            \hline 
            Context Manager & \begin{minipage}{.9\textwidth} \includegraphics[width=15mm]{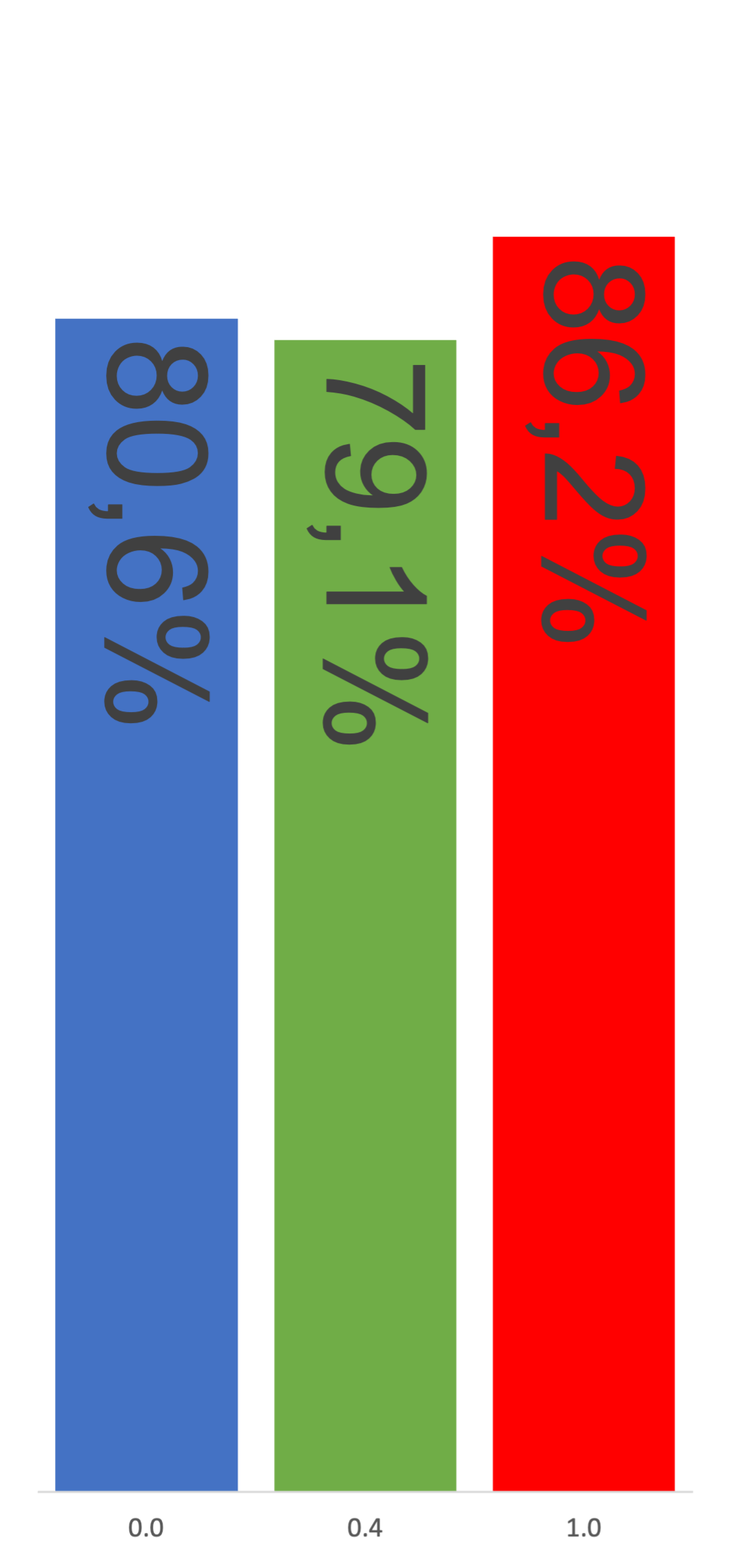} \end{minipage} 
			& \begin{minipage}{.2\textwidth} \includegraphics[width=15mm]{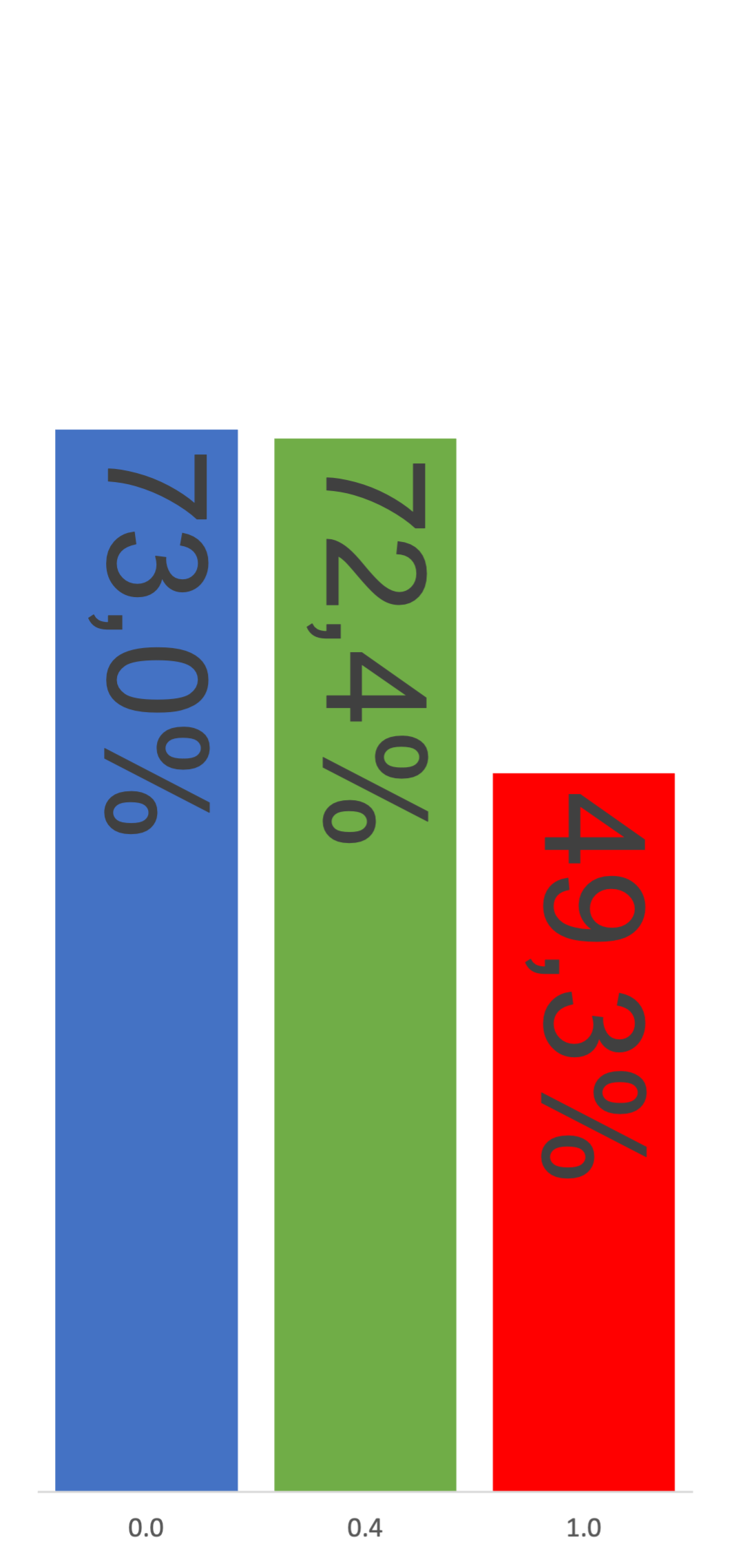} \end{minipage}
			& \begin{minipage}{.2\textwidth} \includegraphics[width=15mm]{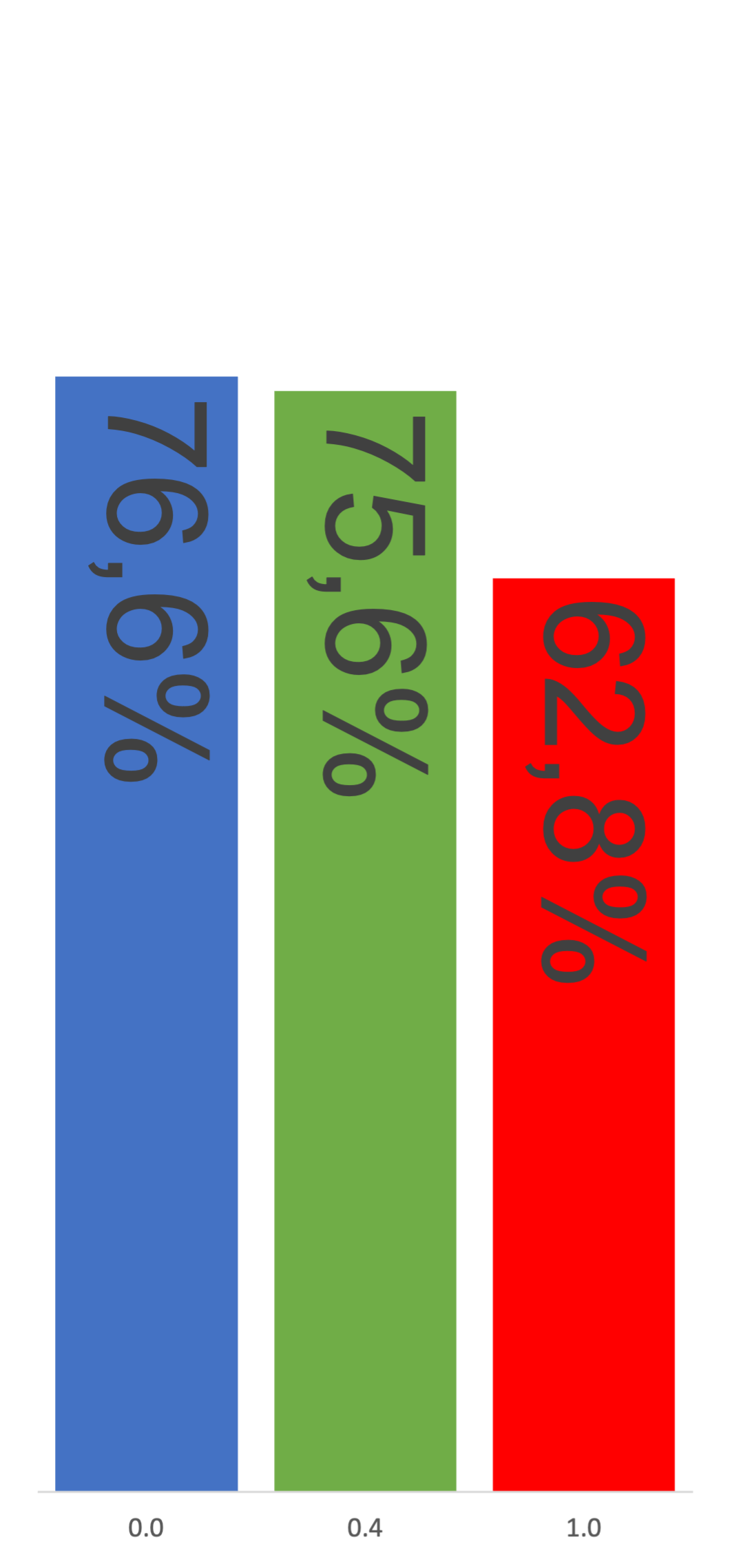} \end{minipage}
			& \begin{minipage}{.2\textwidth} \includegraphics[width=15mm]{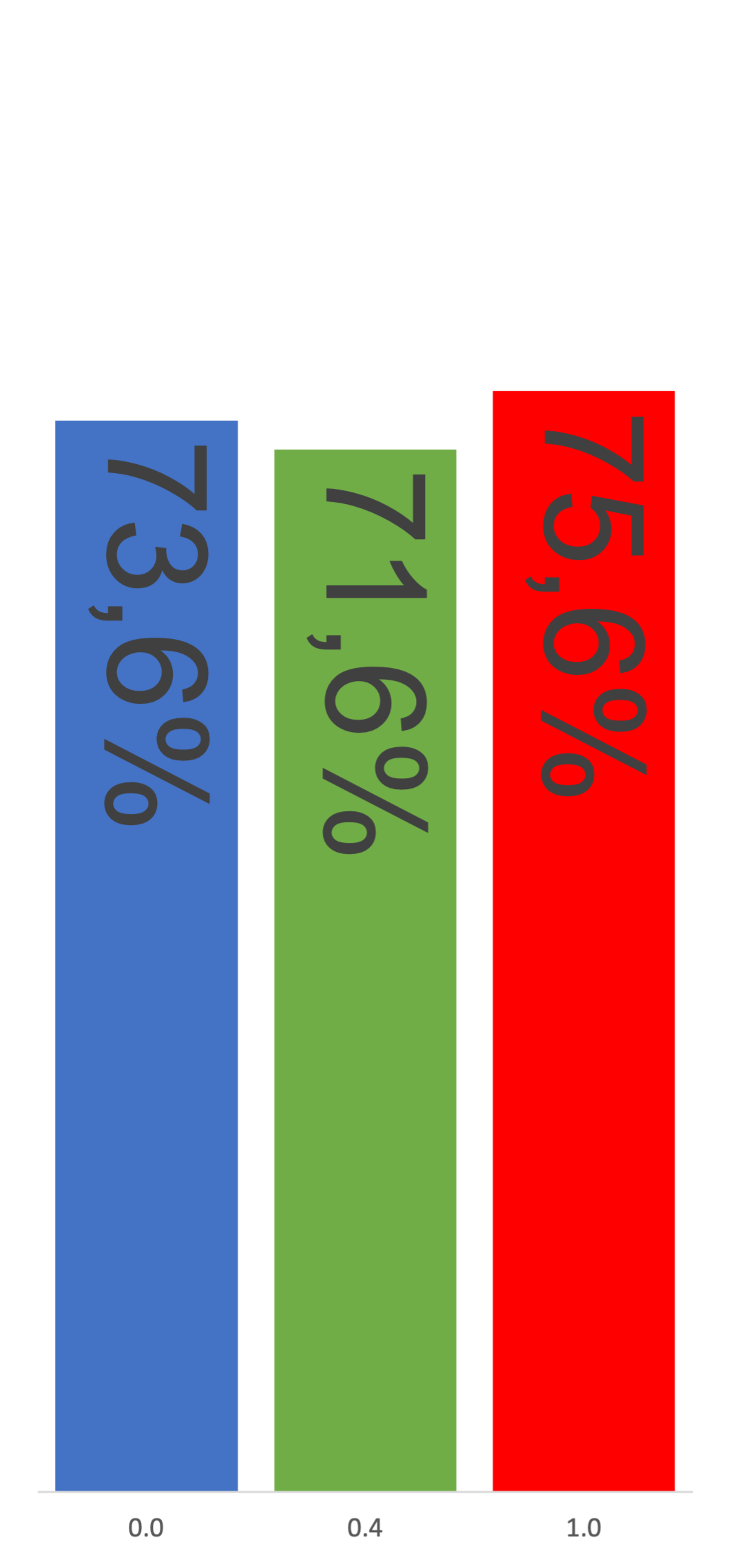} \end{minipage}\\
                \hline
            Persona & \begin{minipage}{.9\textwidth} \includegraphics[width=15mm]{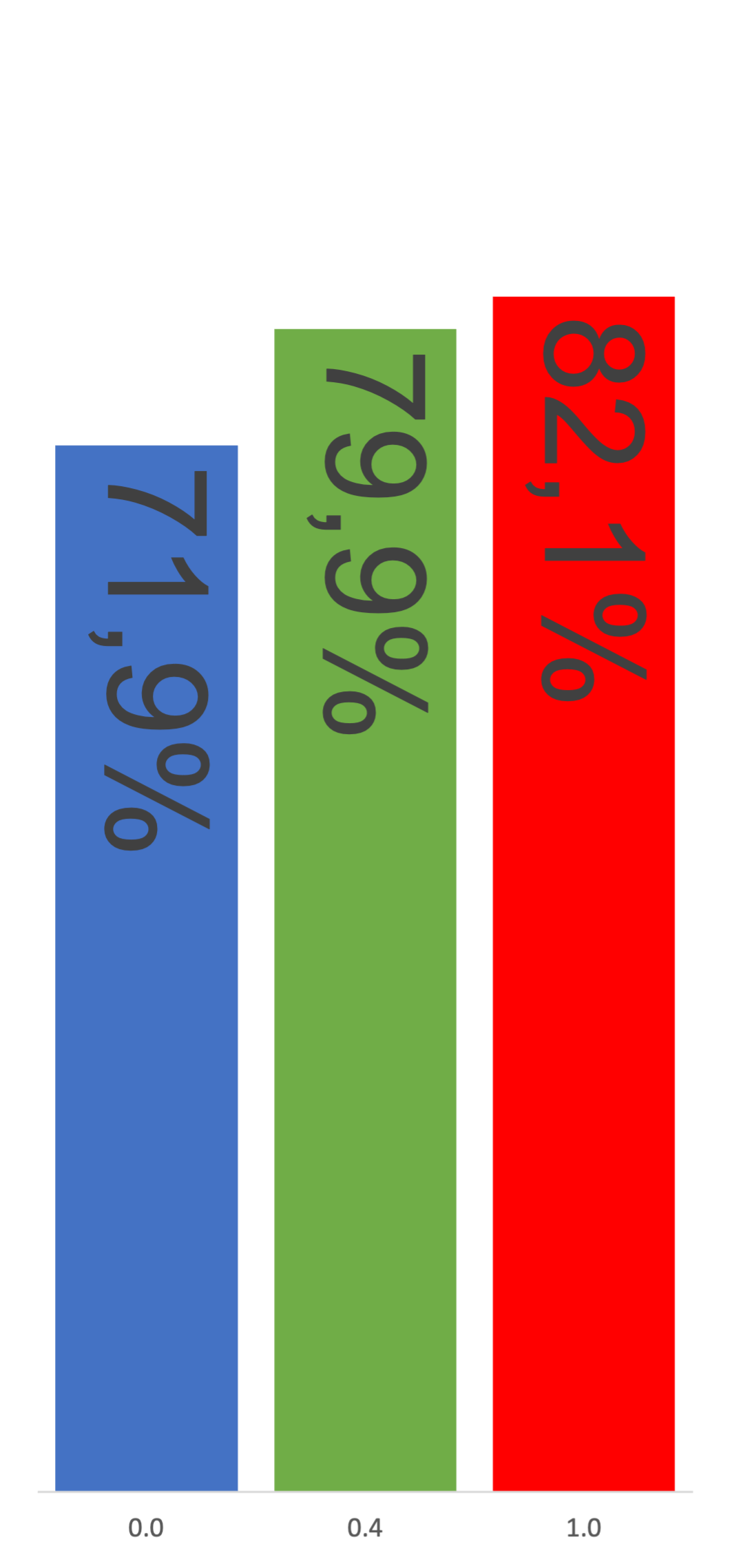} \end{minipage} 
			& \begin{minipage}{.2\textwidth} \includegraphics[width=15mm]{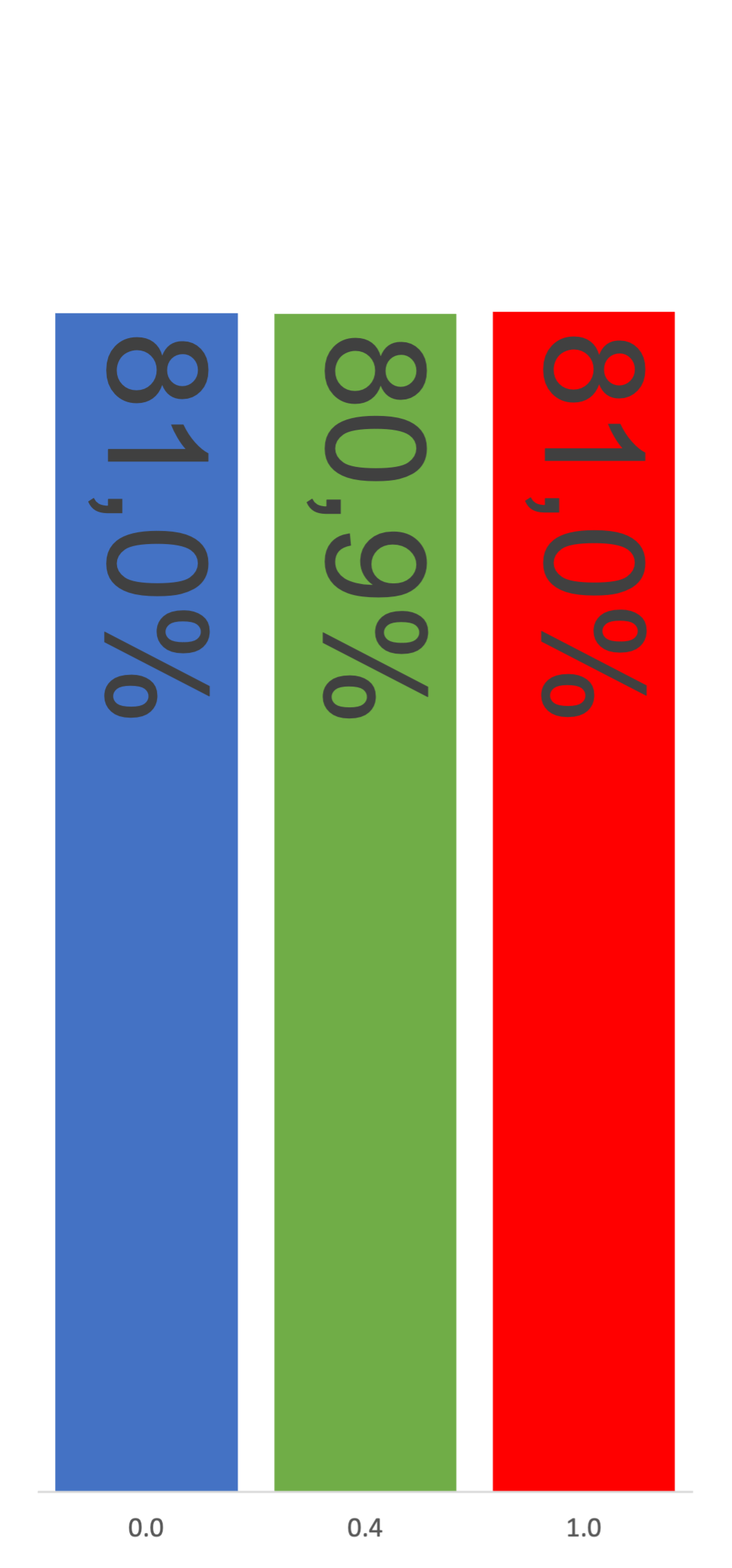} \end{minipage}
			& \begin{minipage}{.2\textwidth} \includegraphics[width=15mm]{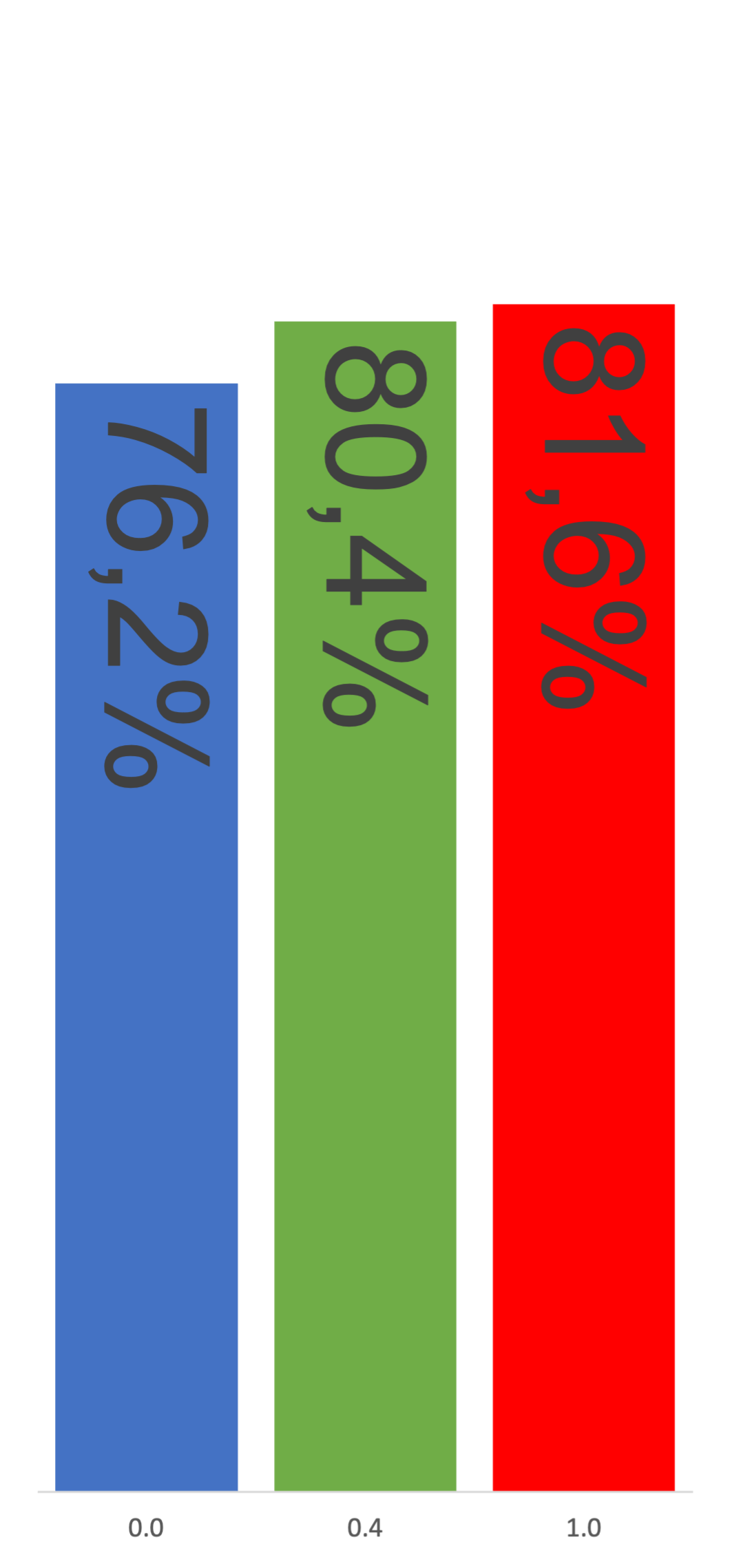} \end{minipage}
			& \begin{minipage}{.2\textwidth} \includegraphics[width=15mm]{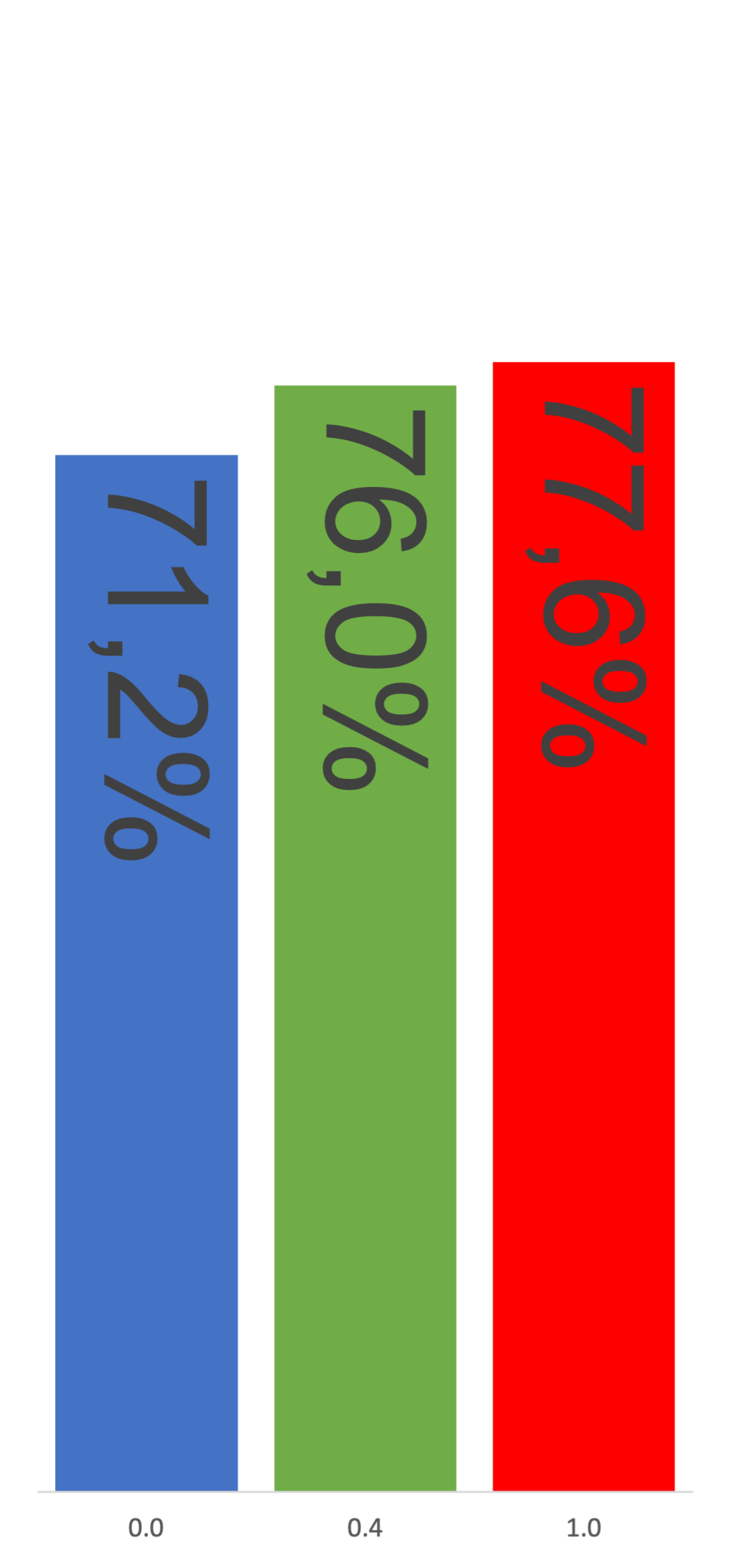} \end{minipage}\\
                \hline
            Question Refinement & \begin{minipage}{.9\textwidth} \includegraphics[width=15mm]{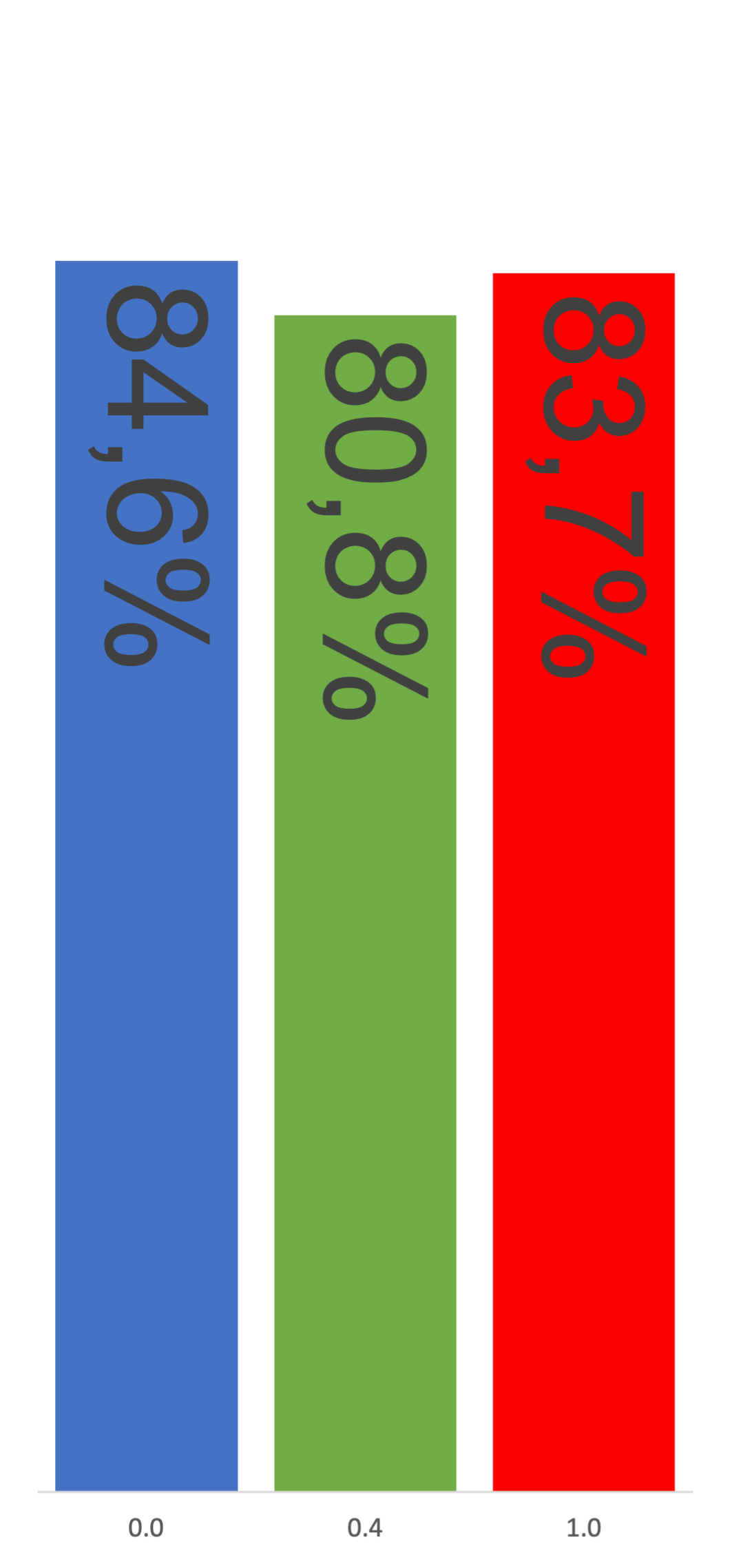} \end{minipage} 
			& \begin{minipage}{.2\textwidth} \includegraphics[width=15mm]{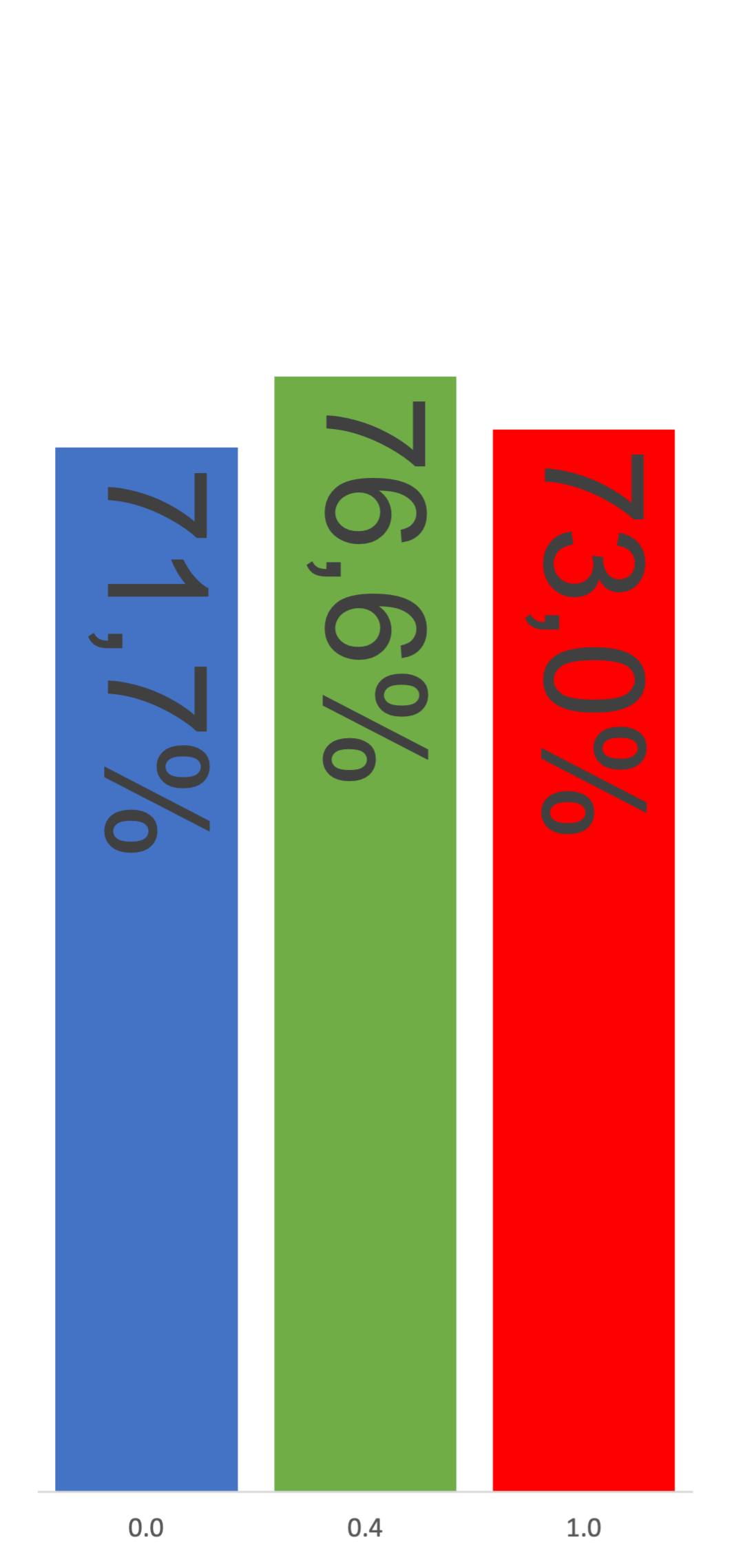} \end{minipage}
			& \begin{minipage}{.2\textwidth} \includegraphics[width=15mm]{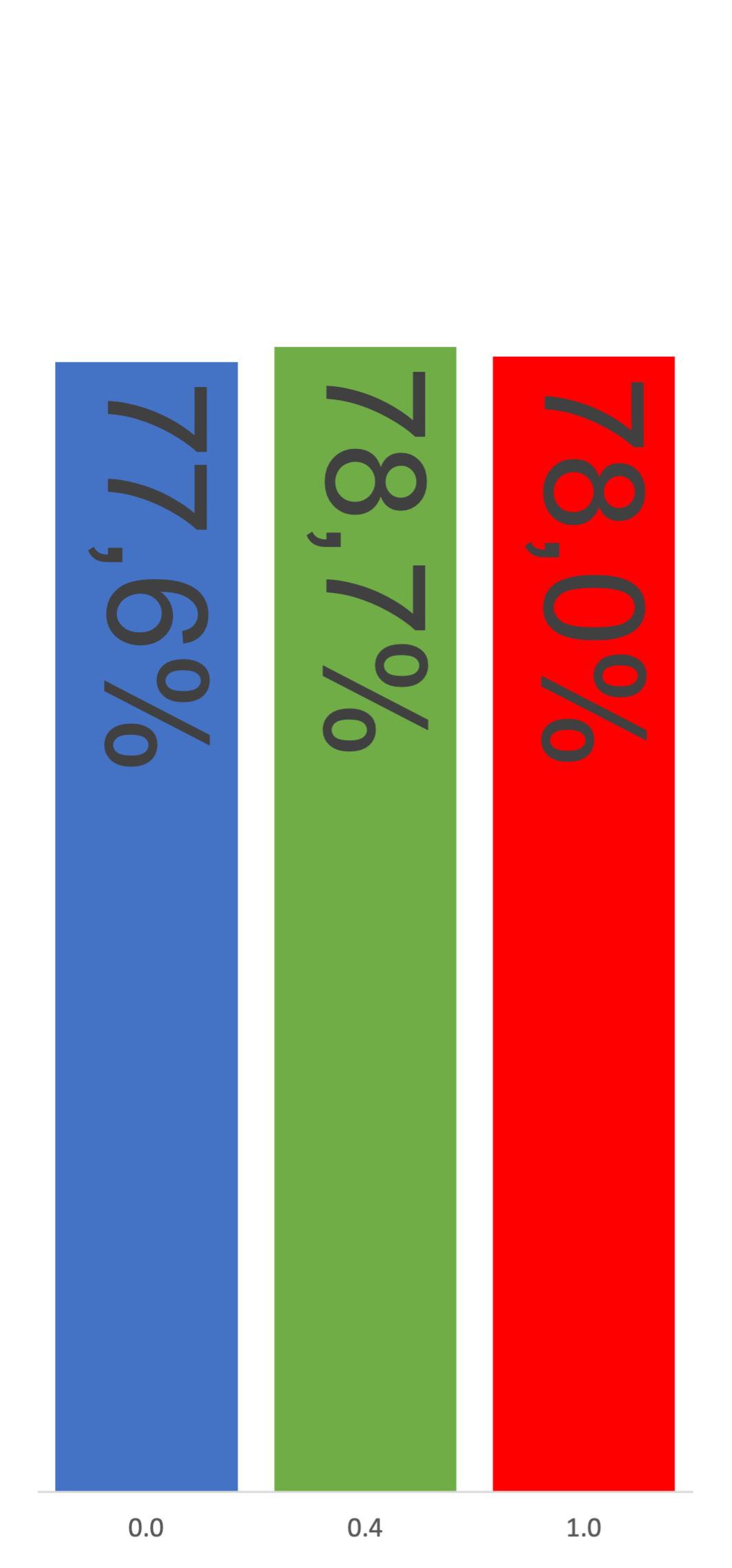} \end{minipage}
			& \begin{minipage}{.2\textwidth} \includegraphics[width=15mm]{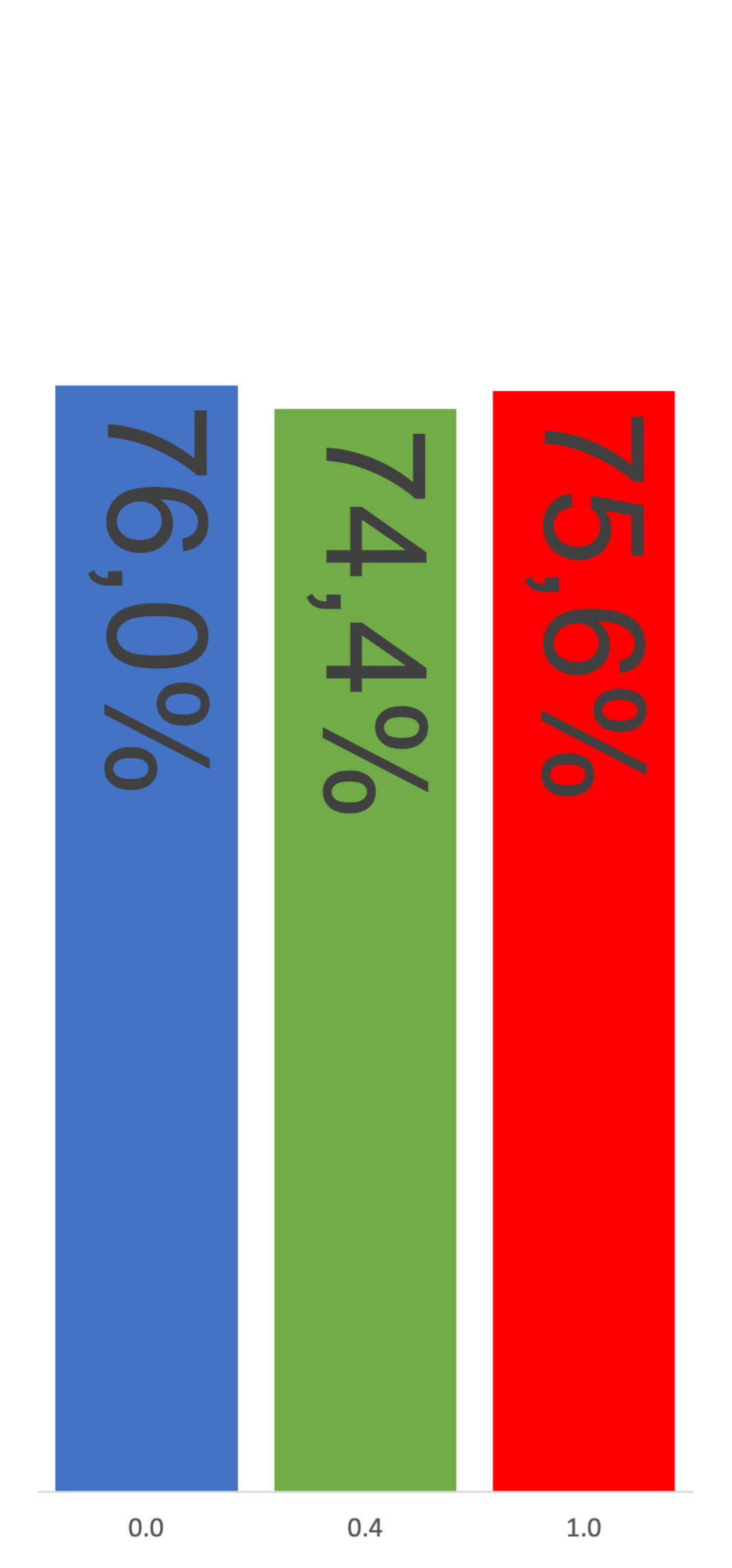} \end{minipage}\\
                \hline
            Template & \begin{minipage}{.9\textwidth} \includegraphics[width=15mm]{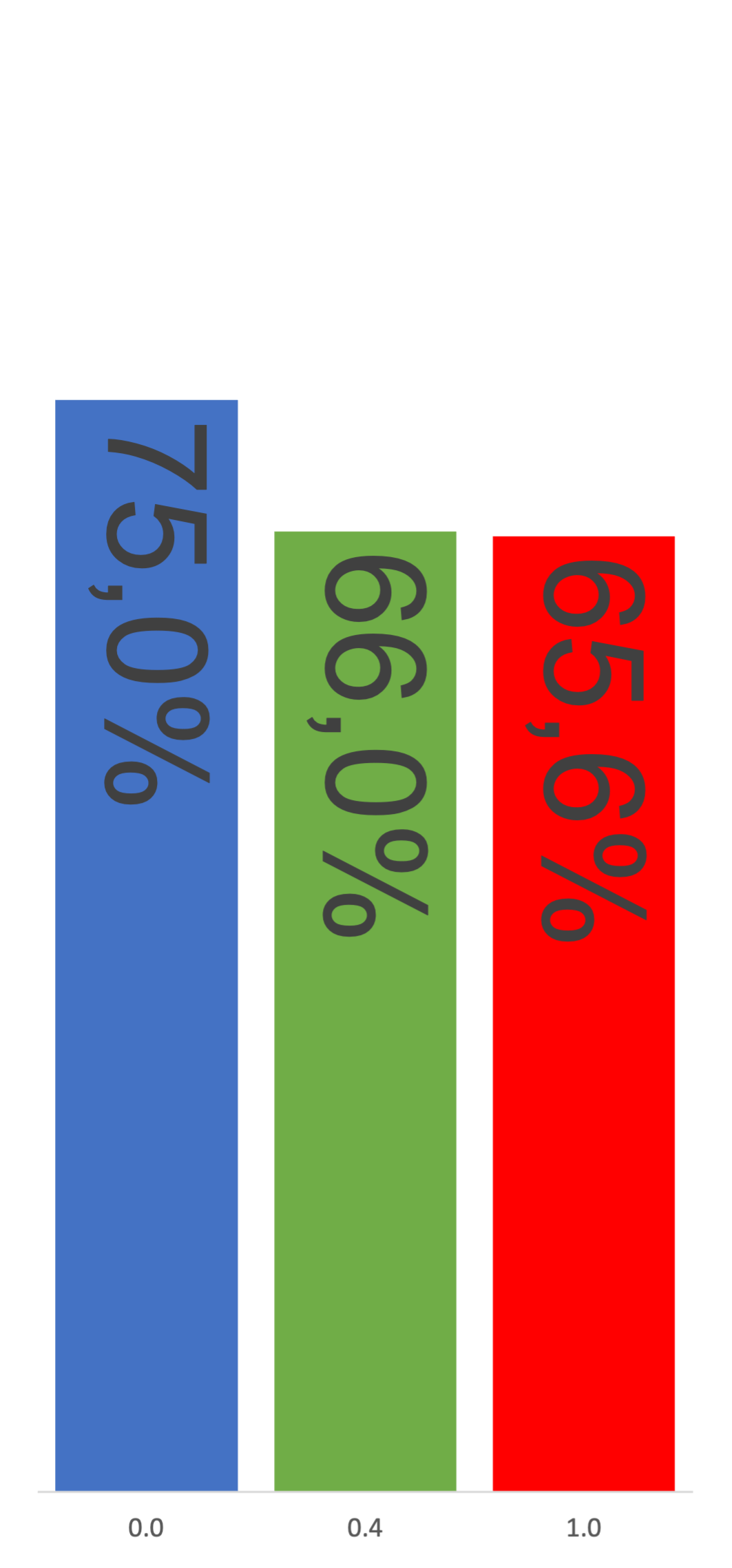} \end{minipage} 
			& \begin{minipage}{.2\textwidth} \includegraphics[width=15mm]{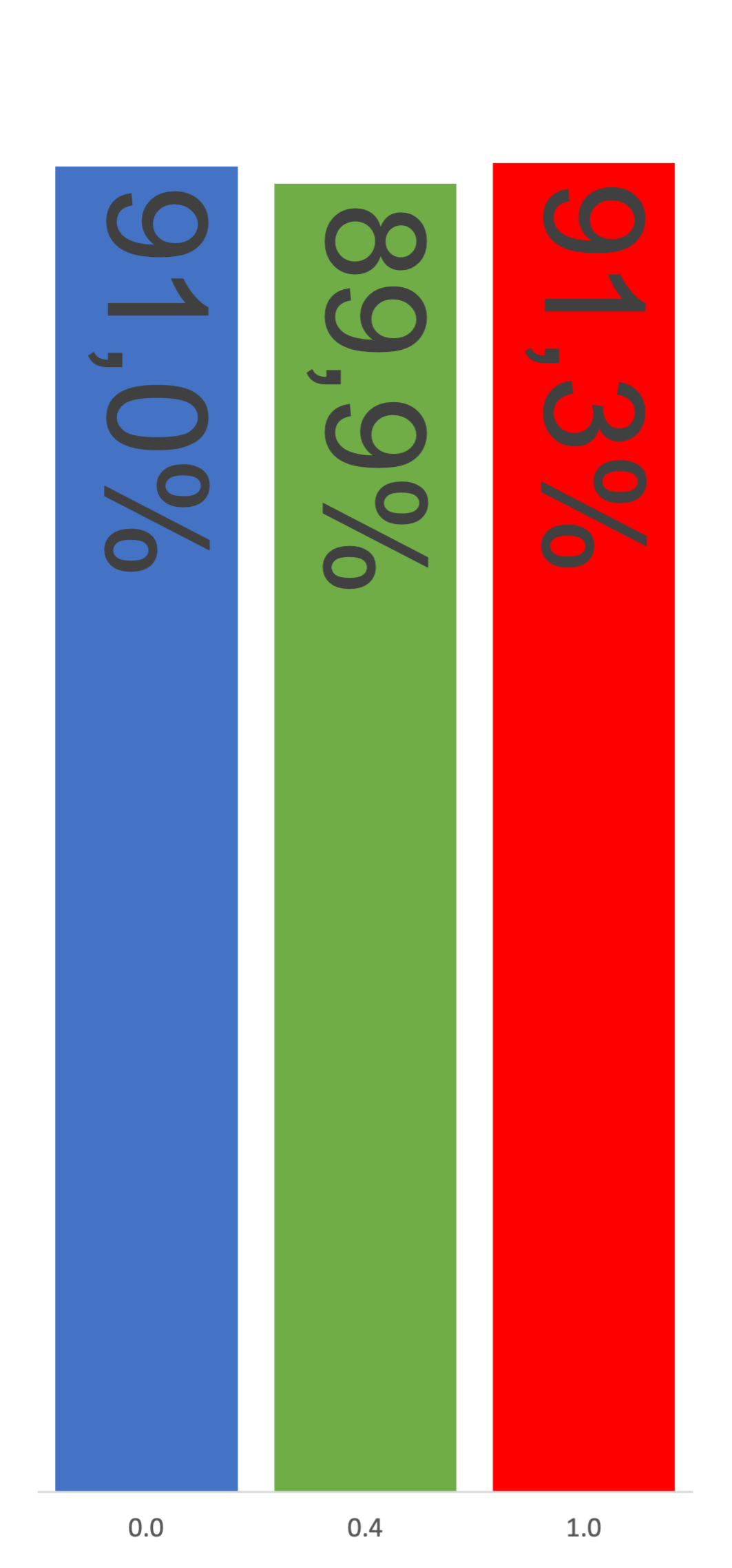} \end{minipage}
			& \begin{minipage}{.2\textwidth} \includegraphics[width=15mm]{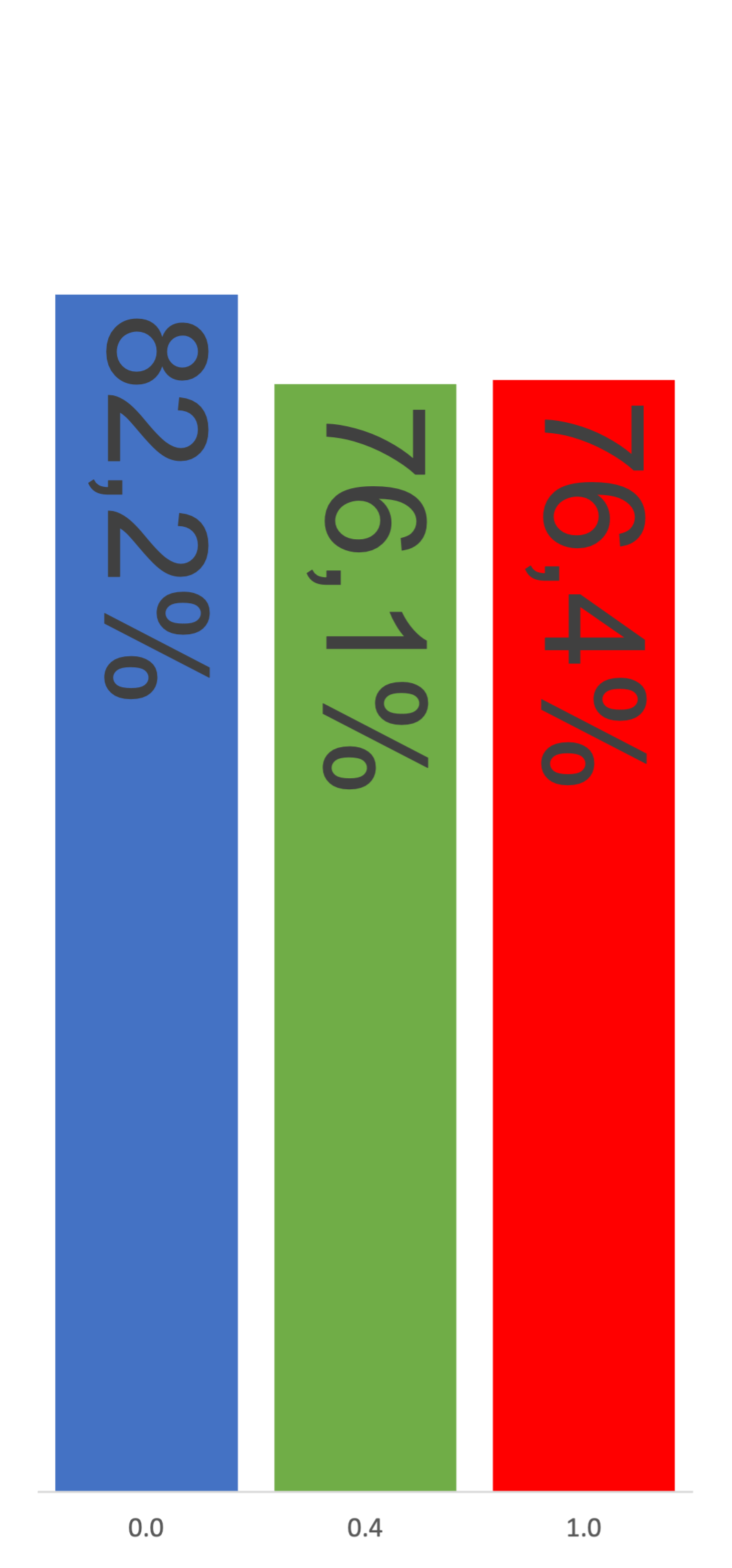} \end{minipage}
			& \begin{minipage}{.2\textwidth} \includegraphics[width=15mm]{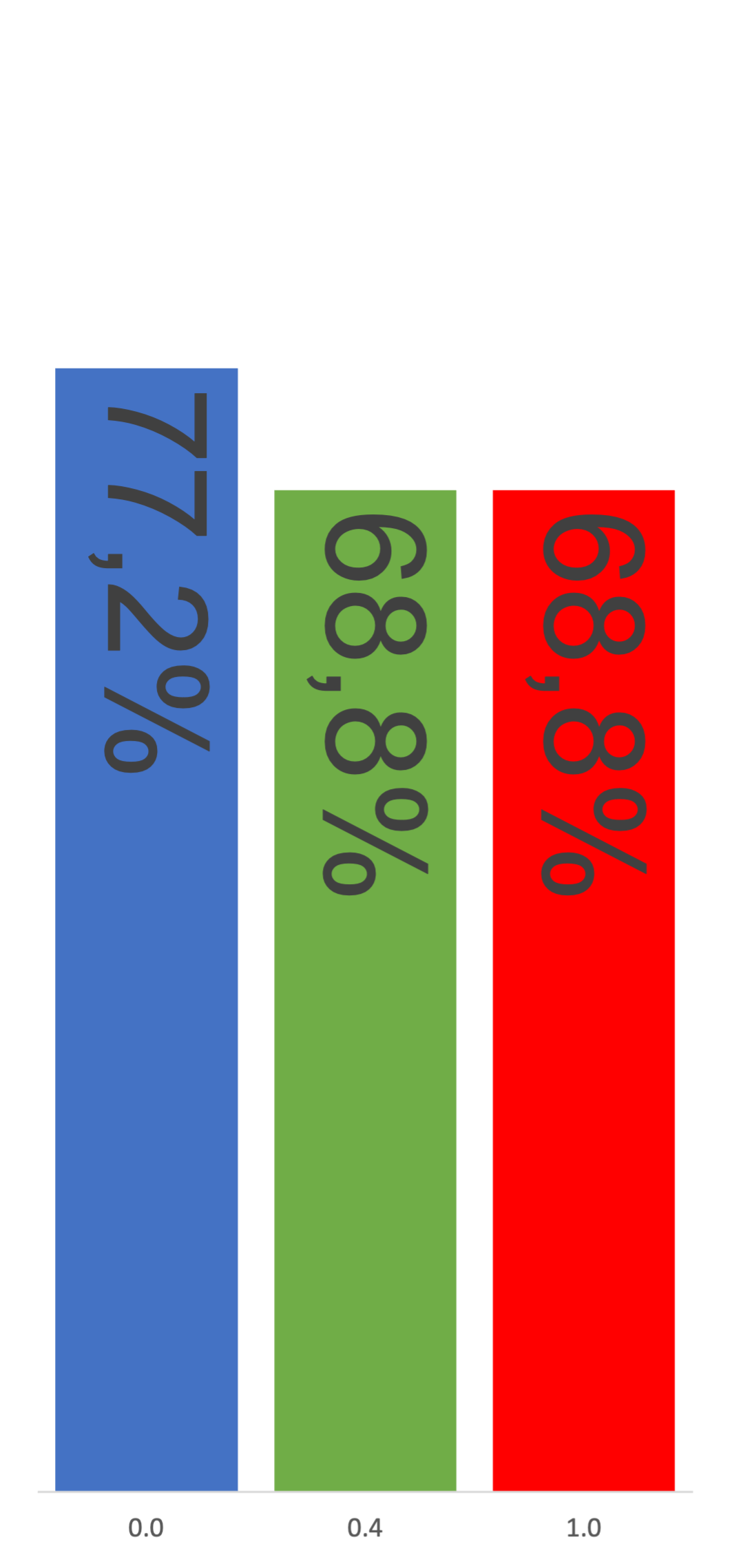} \end{minipage}\\
                \hline    
		\end{tabular}
  }
	\end{threeparttable}
            \caption{Performance measures of the model using all five prompt patterns in binary requirements classification} 
		\label{tab:reqclass}
		\vspace{-.3cm}
\end{table*} 
\raggedbottom

\renewcommand{\arraystretch}{1.3}
\begin{table*}[htbp]
	\centering
	\begin{threeparttable}
 \resizebox{\textwidth}{!}{
		\begin{tabular}{p{0.3\textwidth}p{0.15\textwidth}p{0.15\textwidth}p{0.15\textwidth}p{0.15\textwidth}}
			\toprule 
			\textbf{Prompt Pattern} & \textbf{Precision} & \textbf{Recall}  & \textbf{F-Score} & \textbf{Accuracy} \\
			\hline 
		Cognitive Verifier & \begin{minipage}{.9\textwidth} \includegraphics[width=15mm]{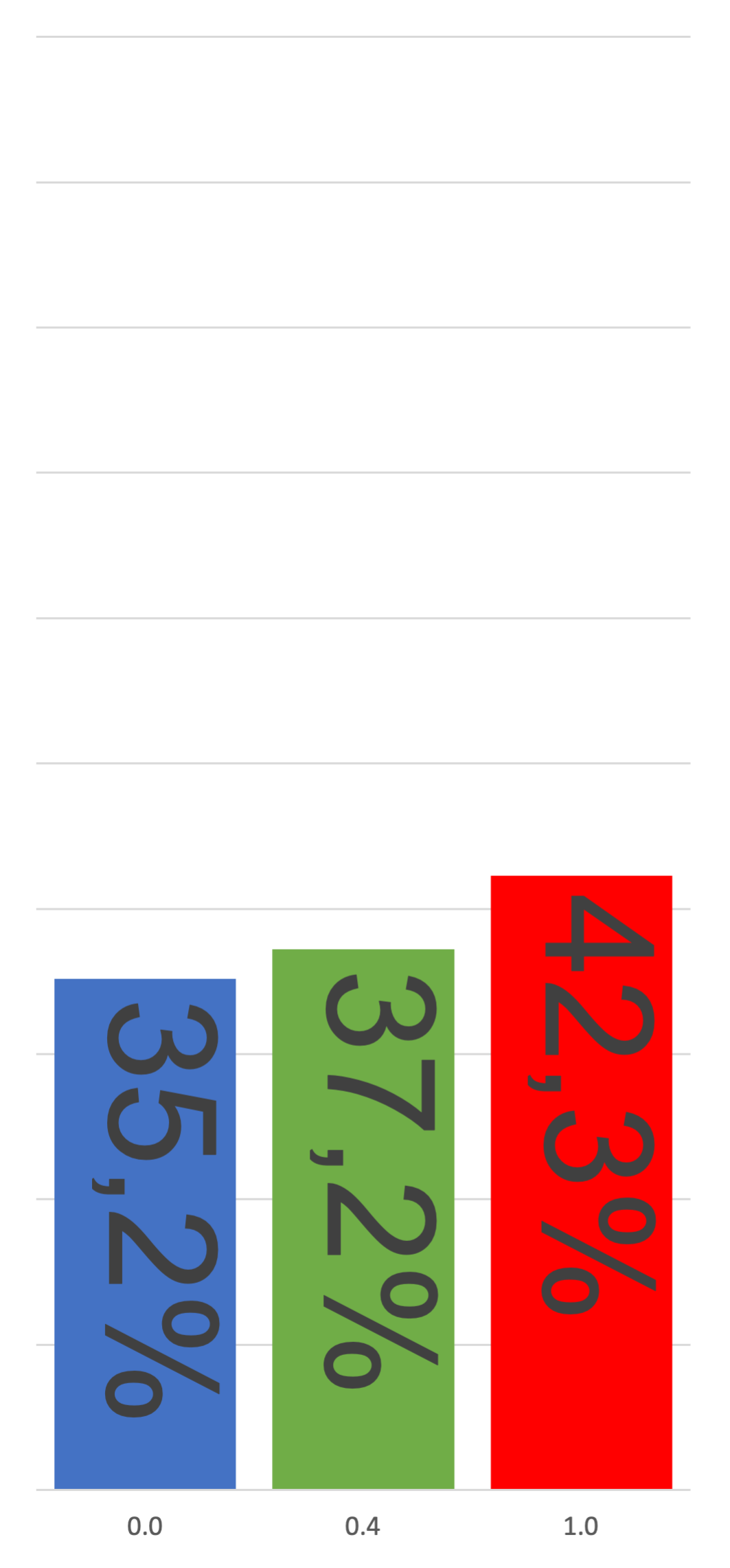} \end{minipage} 
			& \begin{minipage}{.14\textwidth} \includegraphics[width=15mm]{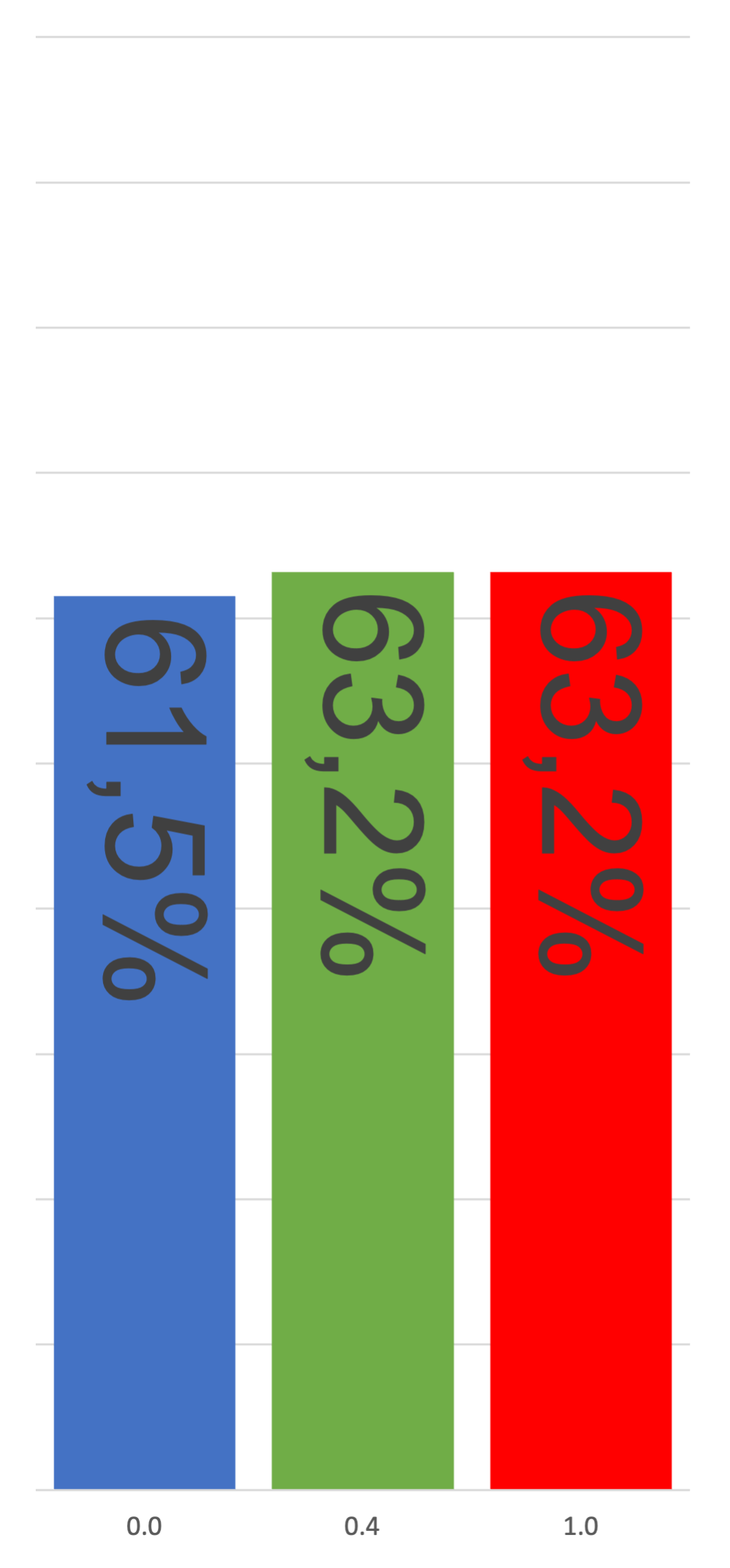} \end{minipage}
			& \begin{minipage}{.14\textwidth} \includegraphics[width=15mm]{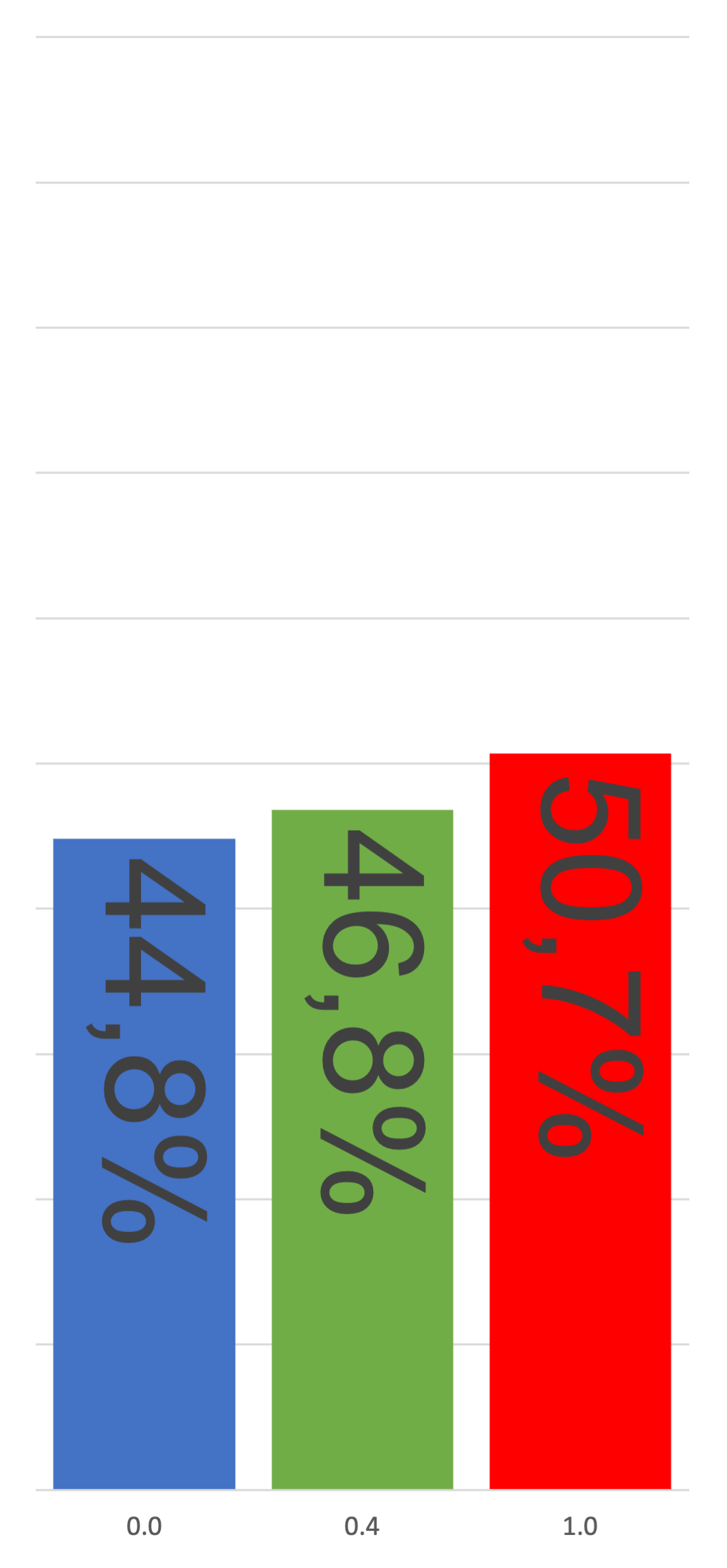} \end{minipage}
			& \begin{minipage}{.14\textwidth} \includegraphics[width=15mm]{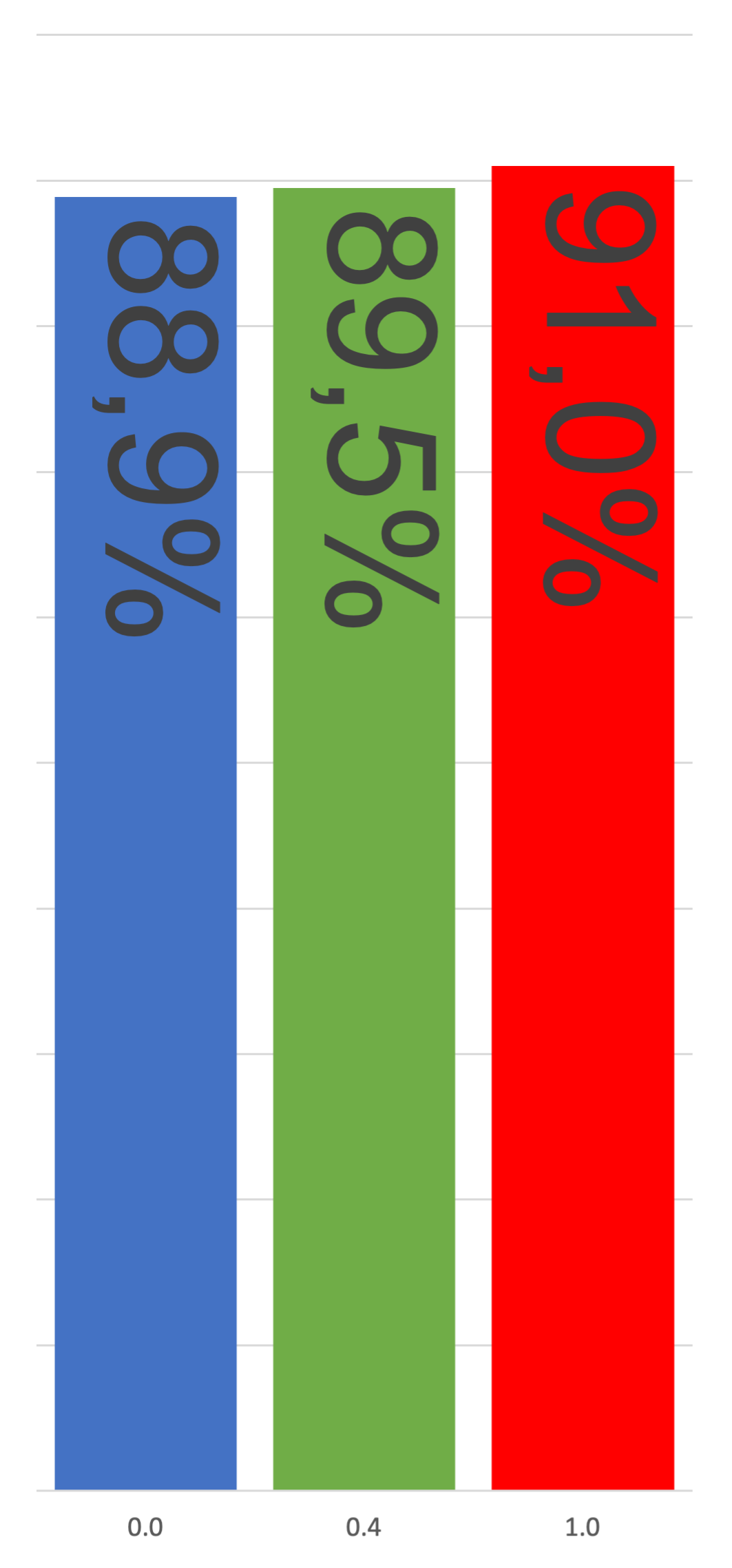} \end{minipage}\\
            \hline 
            Context Manager & \begin{minipage}{.9\textwidth} \includegraphics[width=15mm]{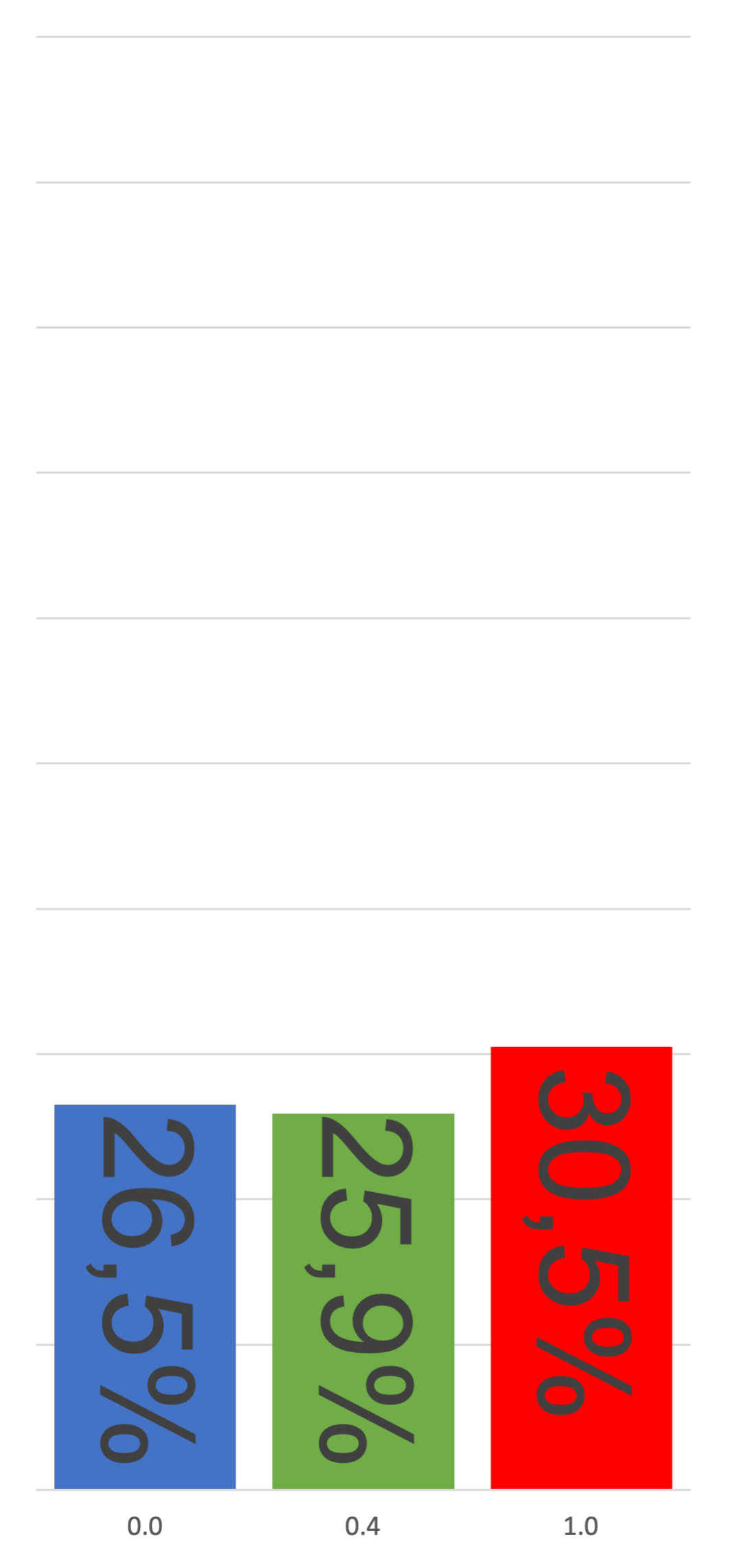} \end{minipage} 
			& \begin{minipage}{.14\textwidth} \includegraphics[width=15mm]{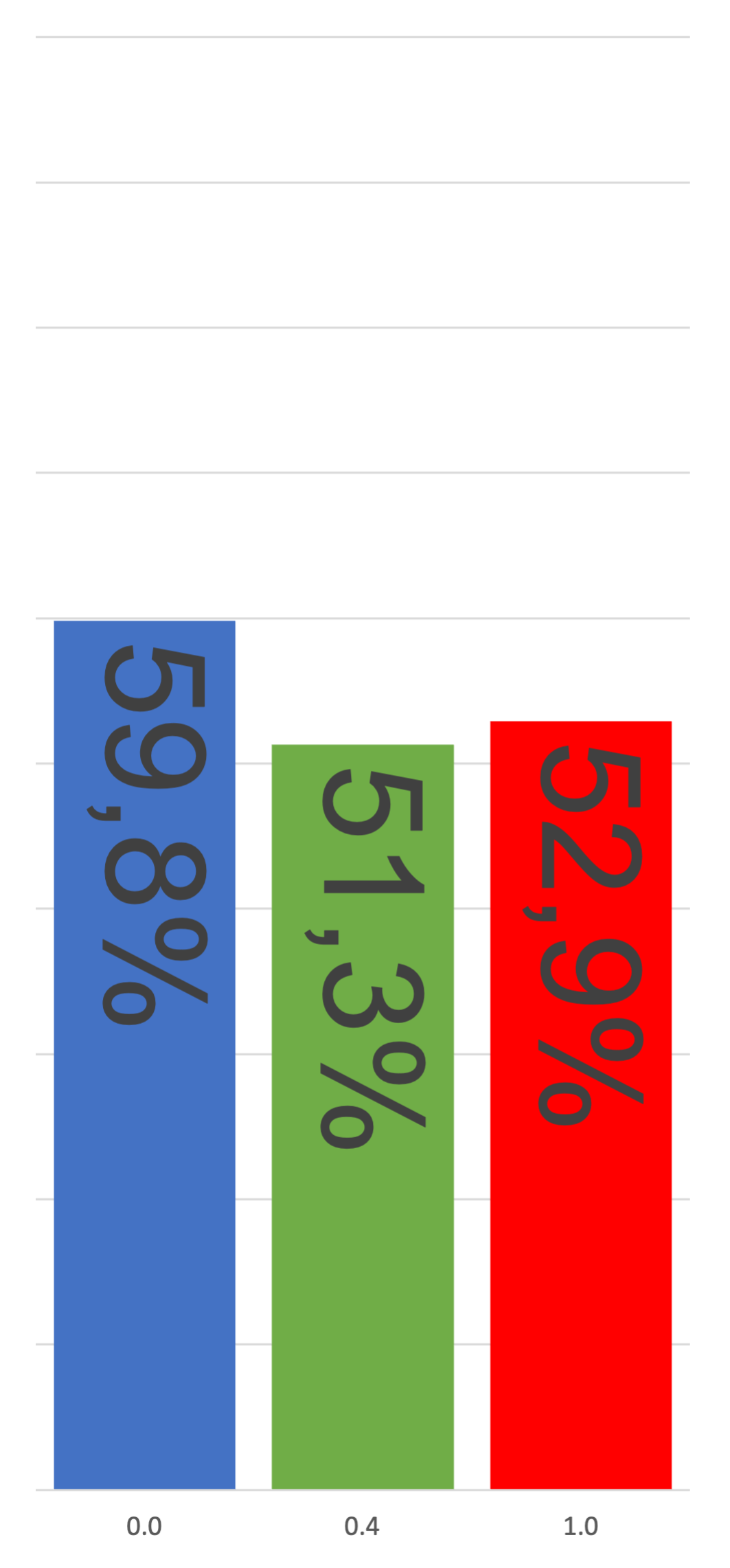} \end{minipage}
			& \begin{minipage}{.14\textwidth} \includegraphics[width=15mm]{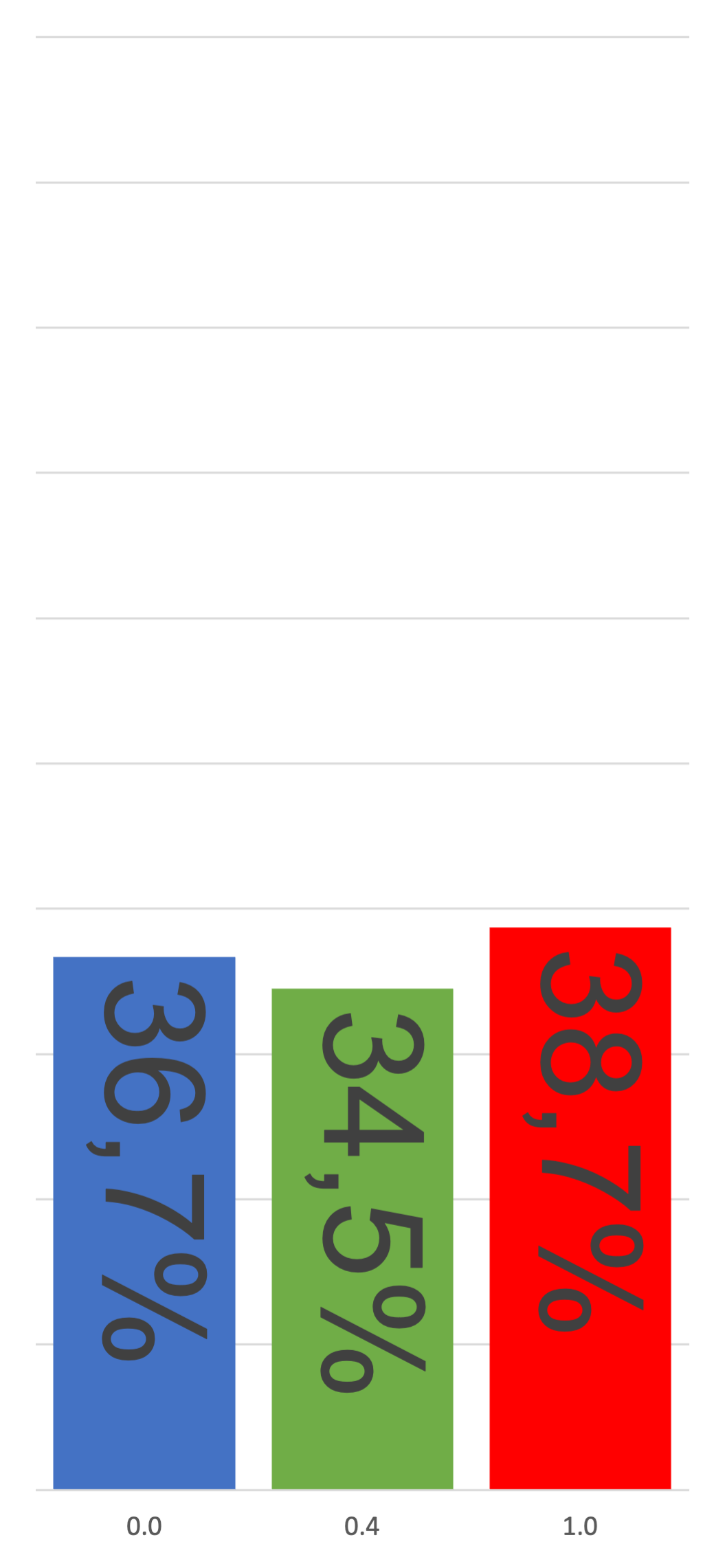} \end{minipage}
			& \begin{minipage}{.14\textwidth} \includegraphics[width=15mm]{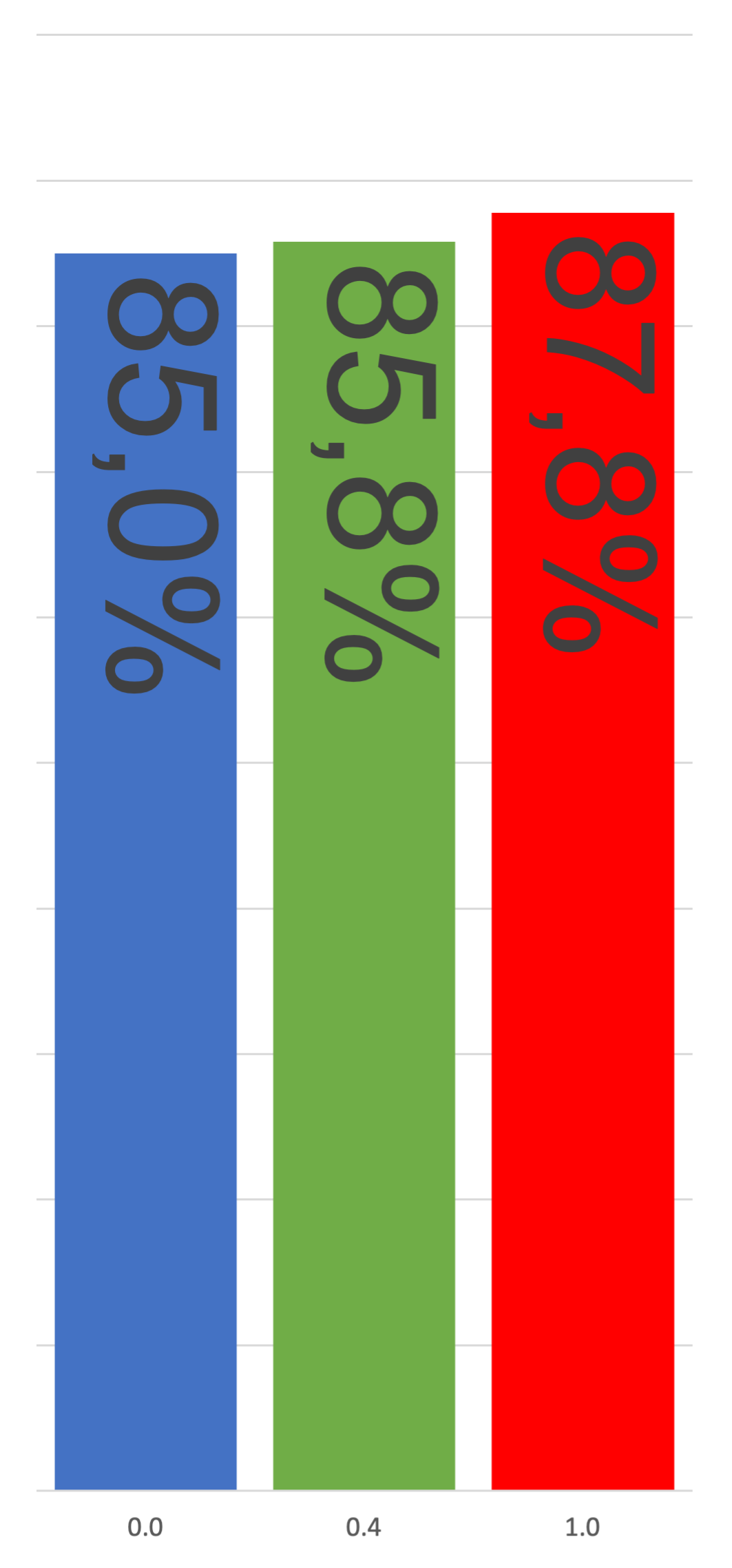} \end{minipage}\\
                \hline
            Persona & \begin{minipage}{.9\textwidth} \includegraphics[width=15mm]{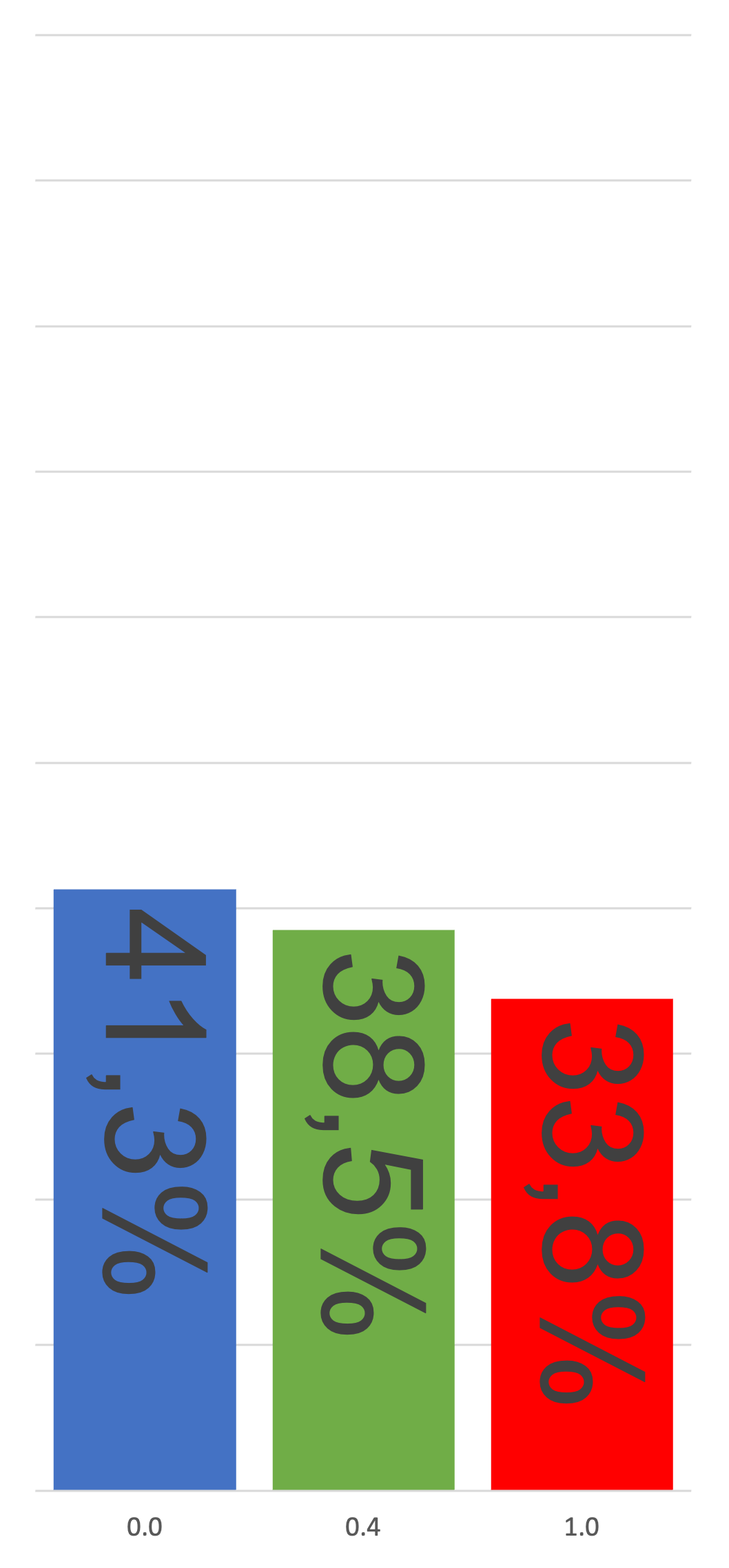} \end{minipage} 
			& \begin{minipage}{.14\textwidth} \includegraphics[width=15mm]{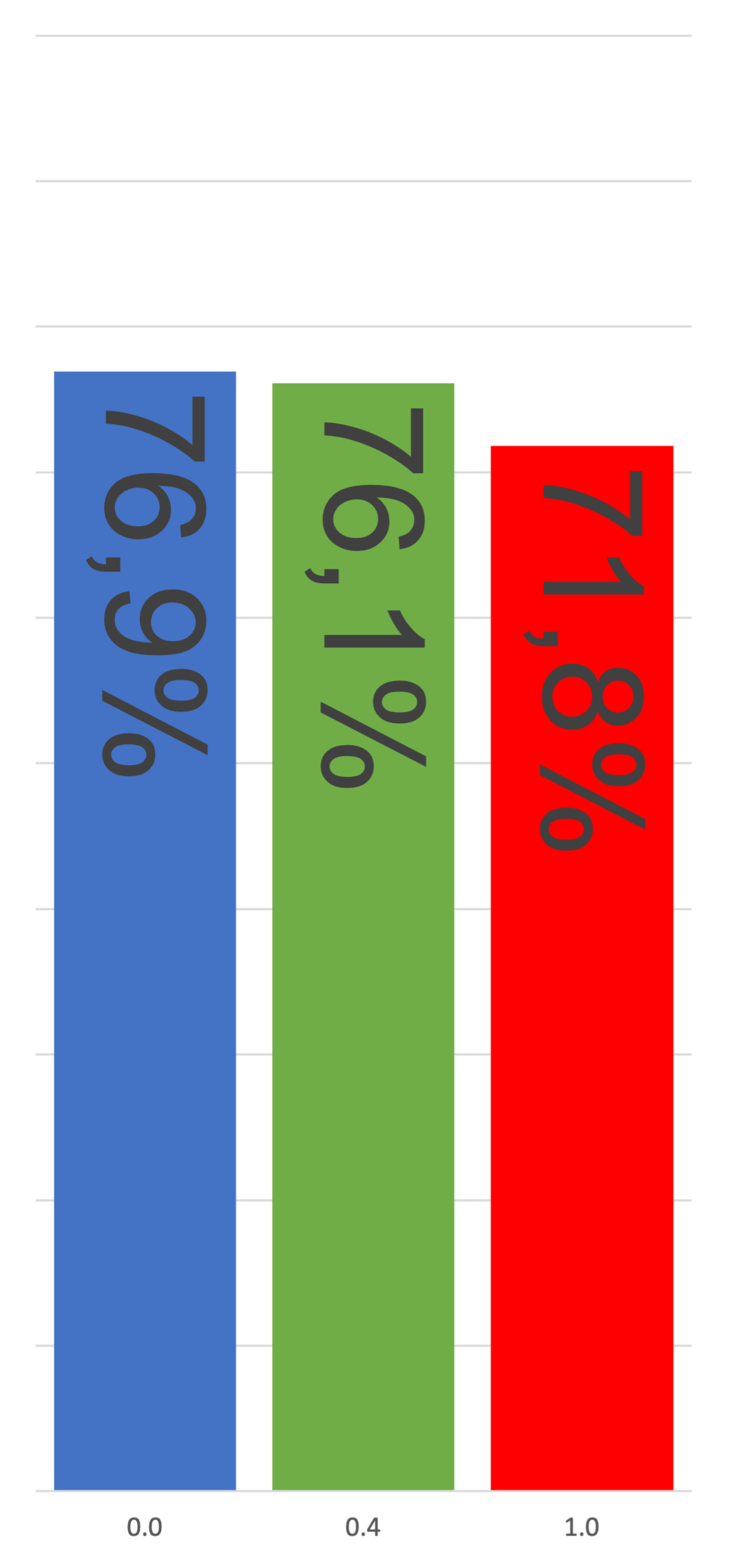} \end{minipage}
			& \begin{minipage}{.14\textwidth} \includegraphics[width=15mm]{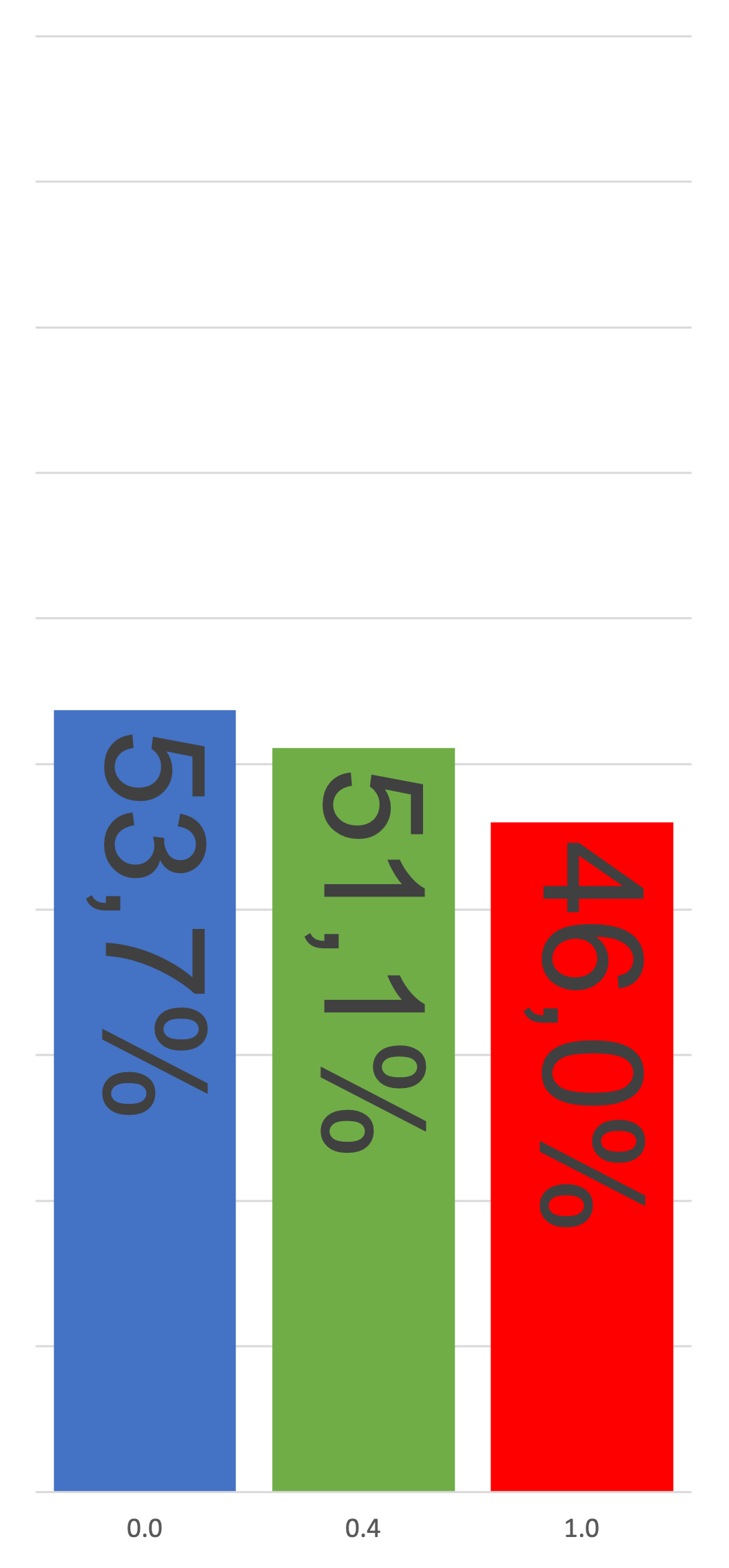} \end{minipage}
			& \begin{minipage}{.14\textwidth} \includegraphics[width=15mm]{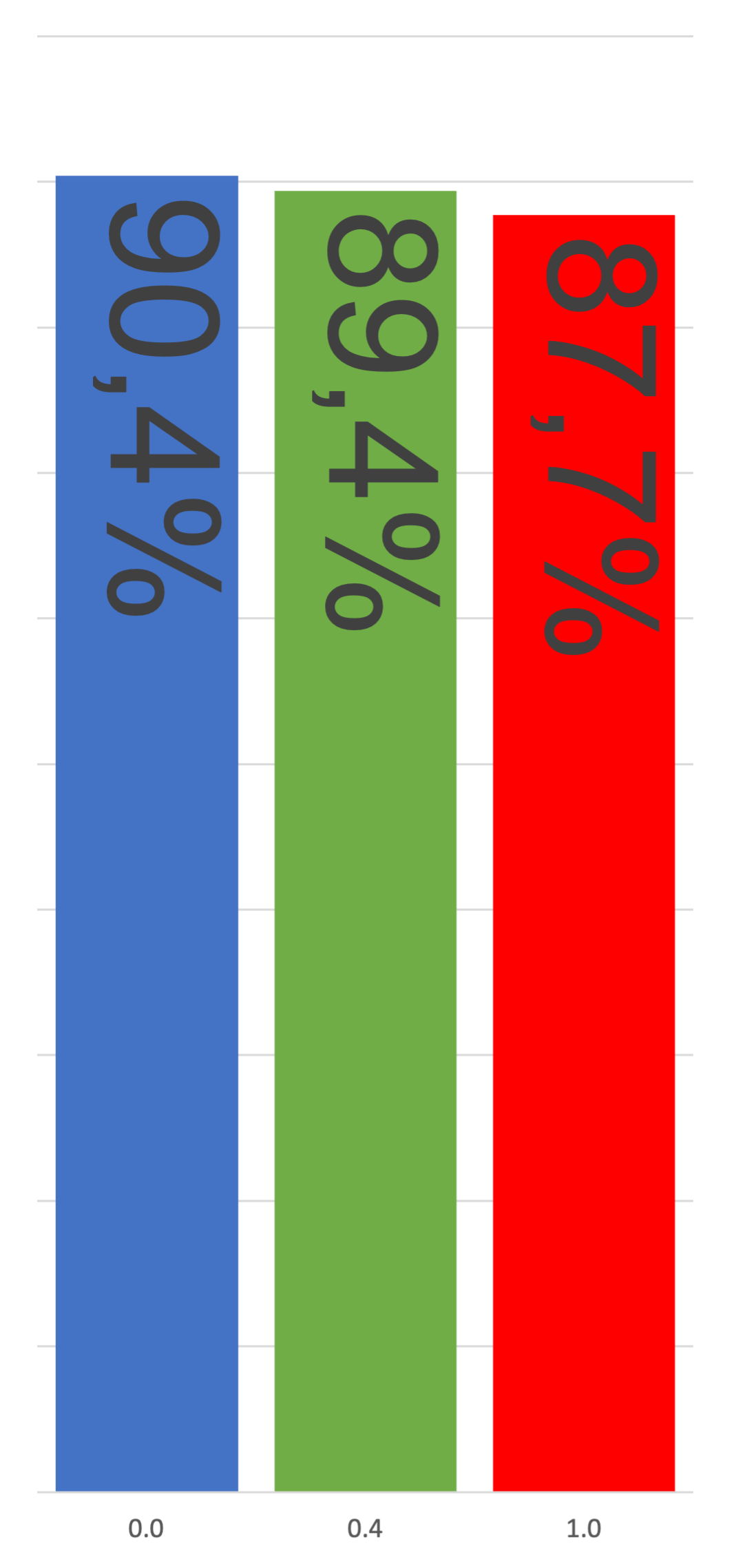} \end{minipage}\\
                \hline
            Question Refinement & \begin{minipage}{.9\textwidth} \includegraphics[width=15mm]{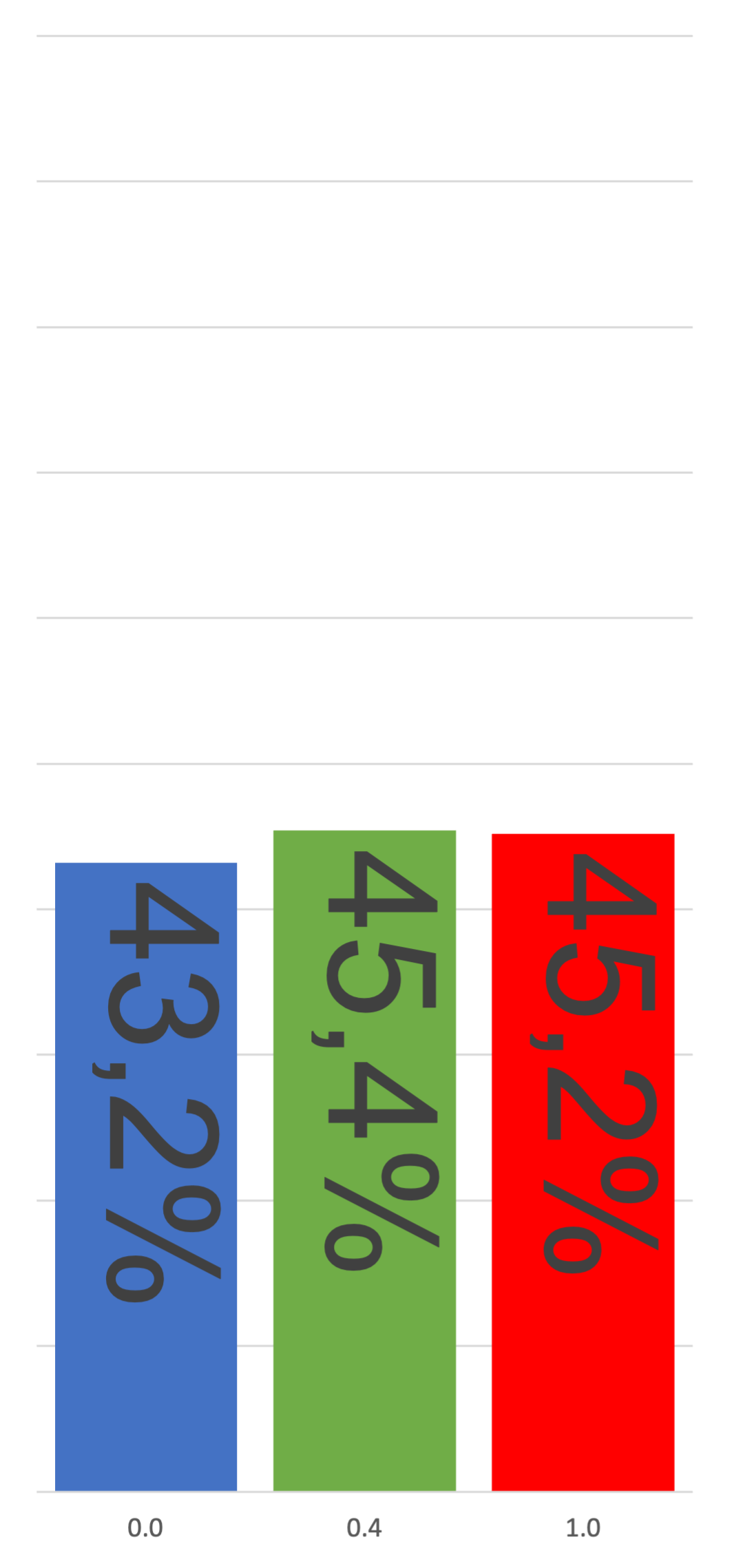} \end{minipage} 
			& \begin{minipage}{.14\textwidth} \includegraphics[width=15mm]{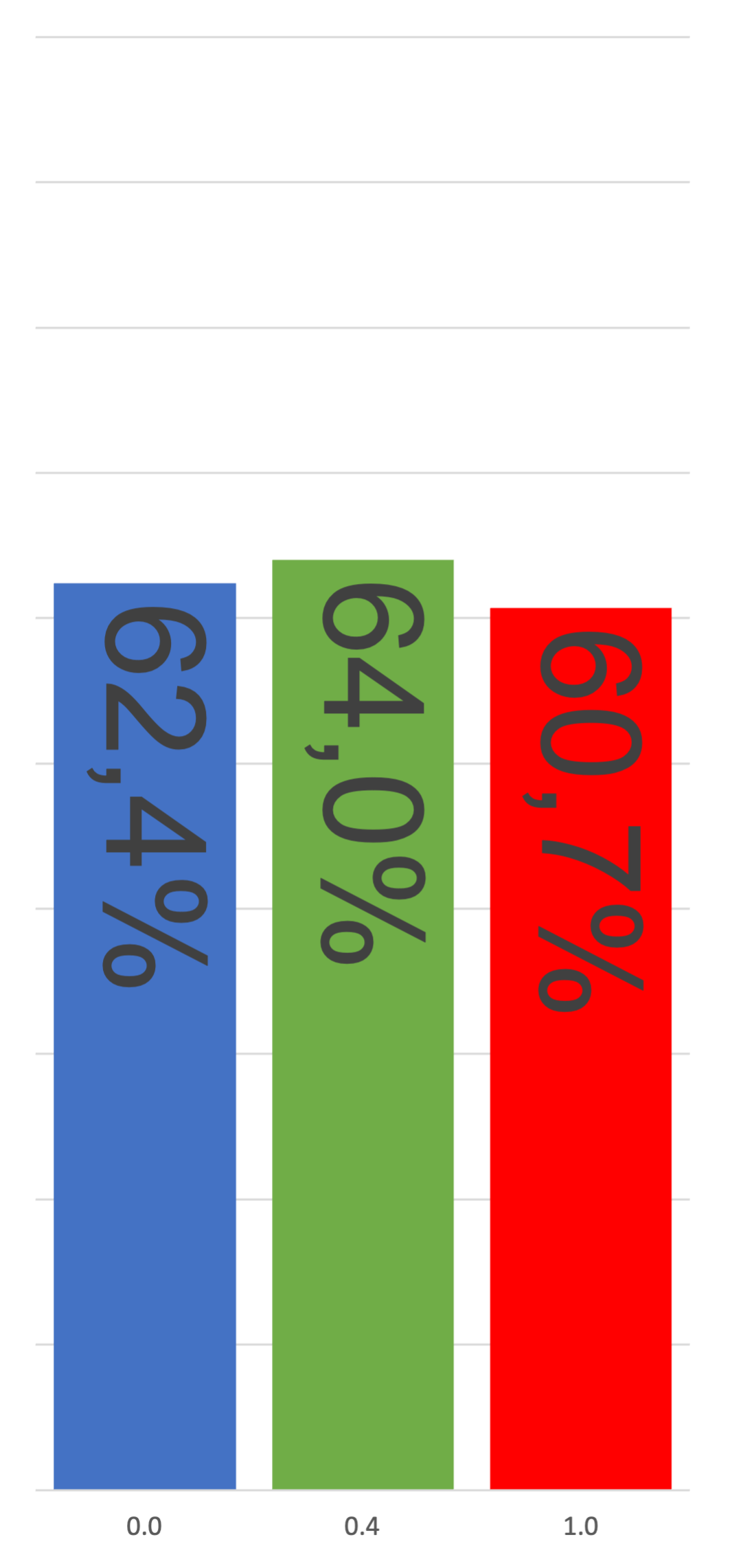} \end{minipage}
			& \begin{minipage}{.14\textwidth} \includegraphics[width=15mm]{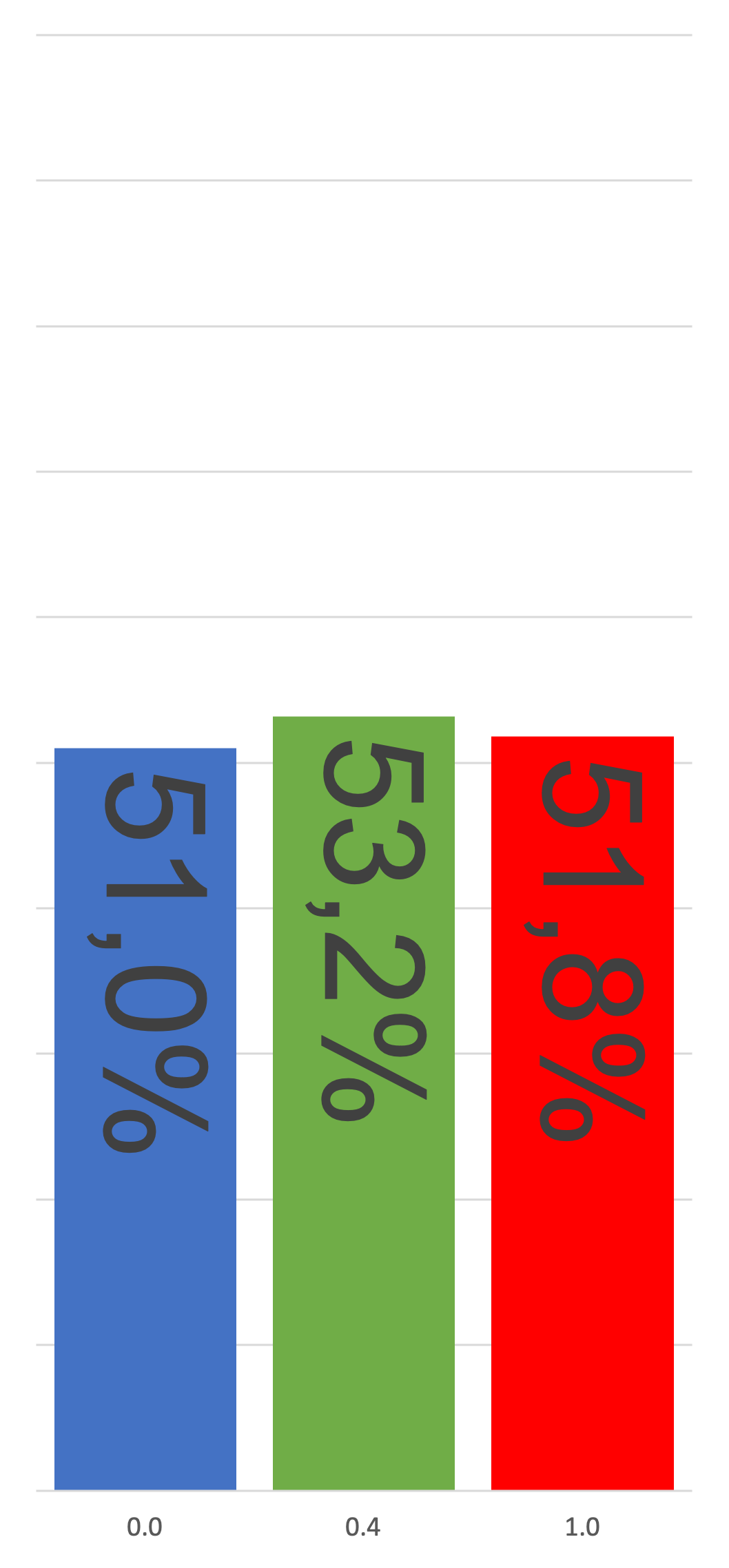} \end{minipage}
			& \begin{minipage}{.14\textwidth} \includegraphics[width=15mm]{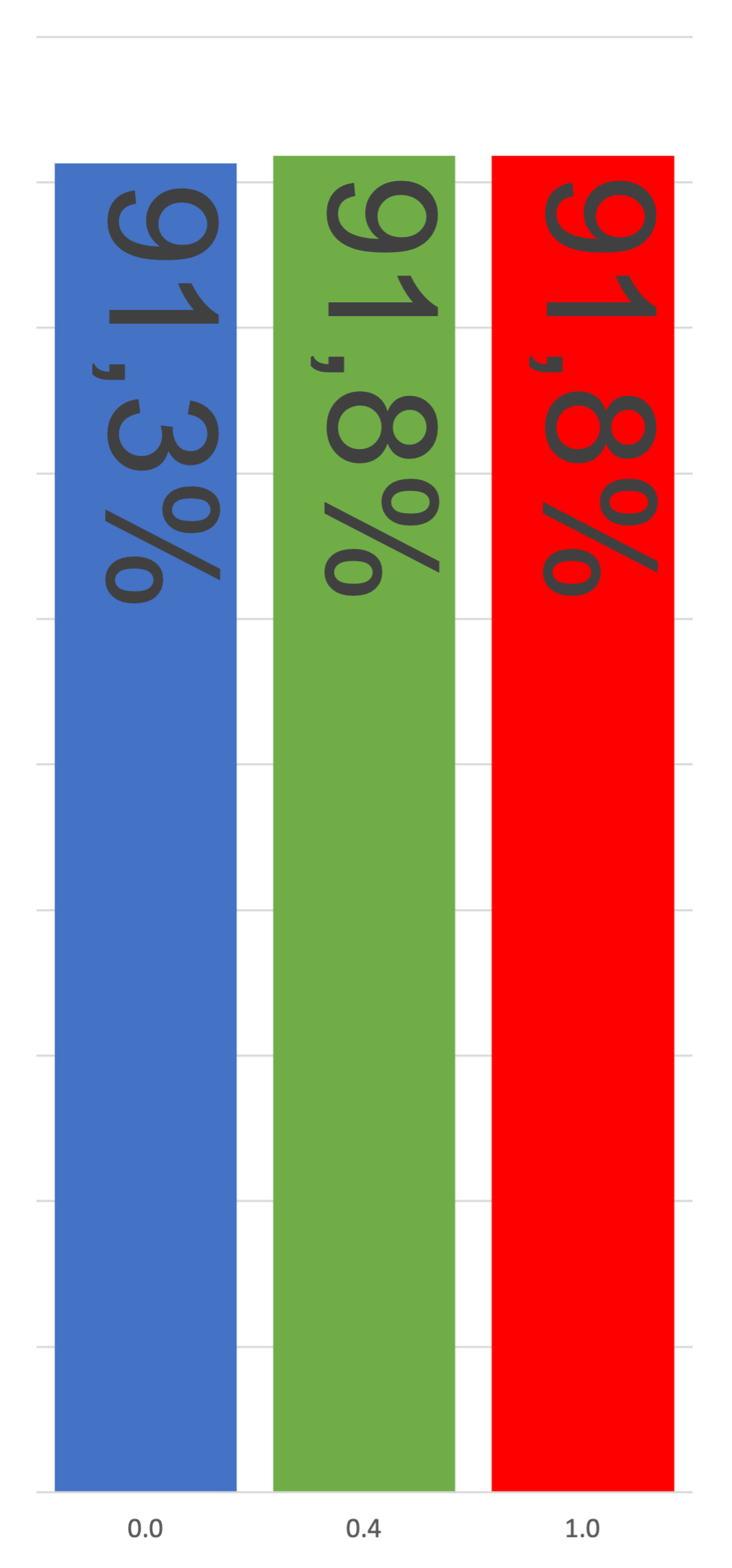} \end{minipage}\\
                \hline
            Template & \begin{minipage}{.9\textwidth} \includegraphics[width=15mm]{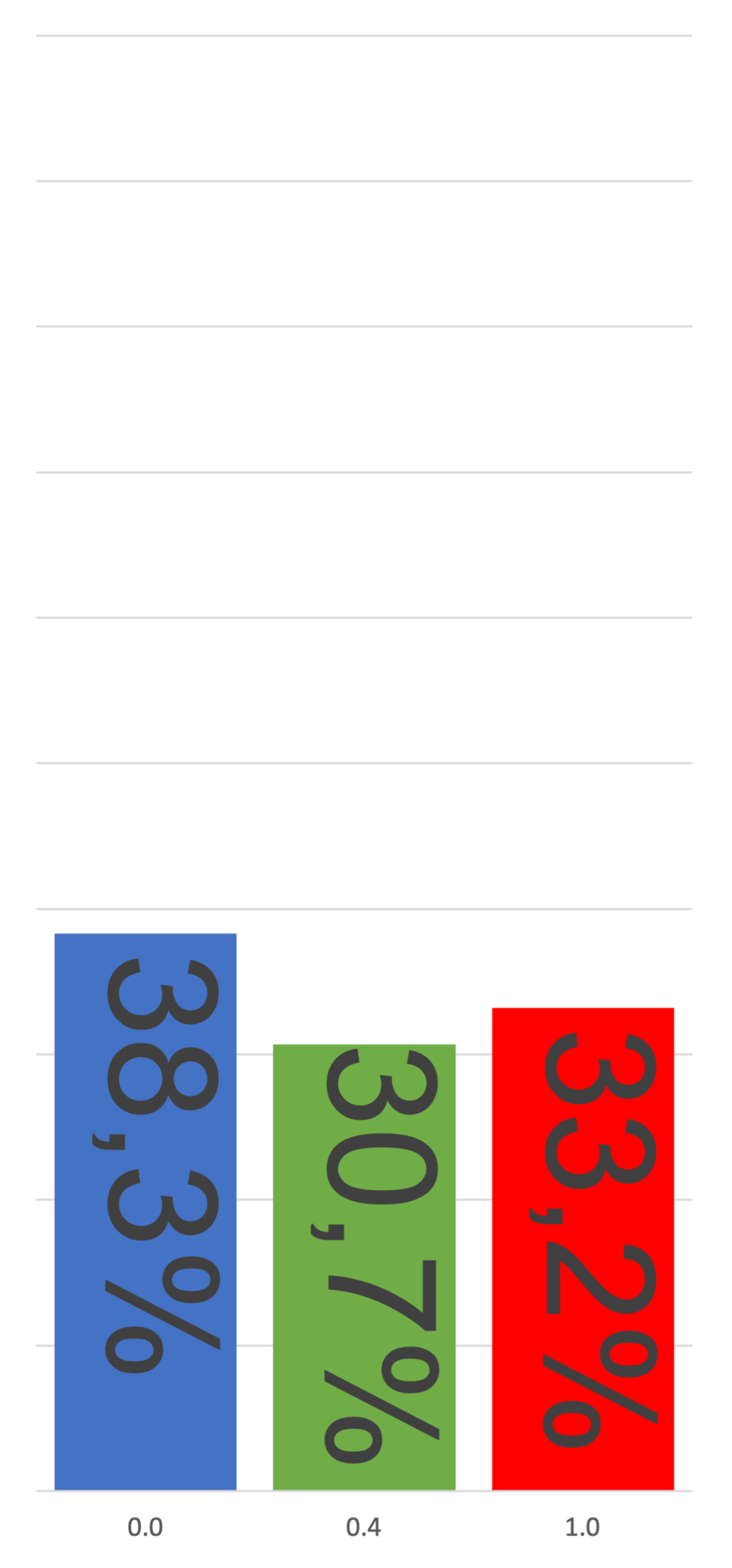} \end{minipage} 
			& \begin{minipage}{.14\textwidth} \includegraphics[width=15mm]{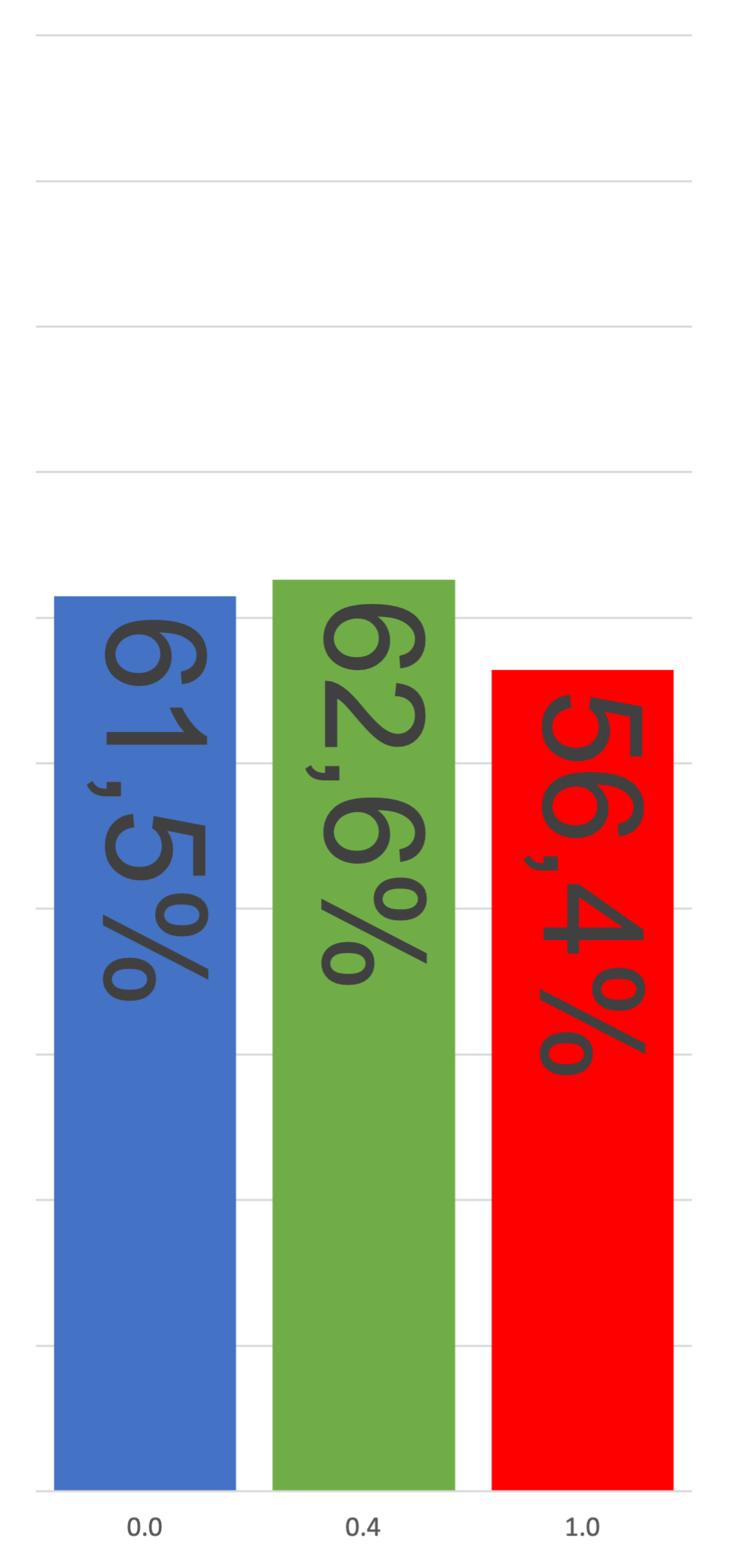} \end{minipage}
			& \begin{minipage}{.14\textwidth} \includegraphics[width=15mm]{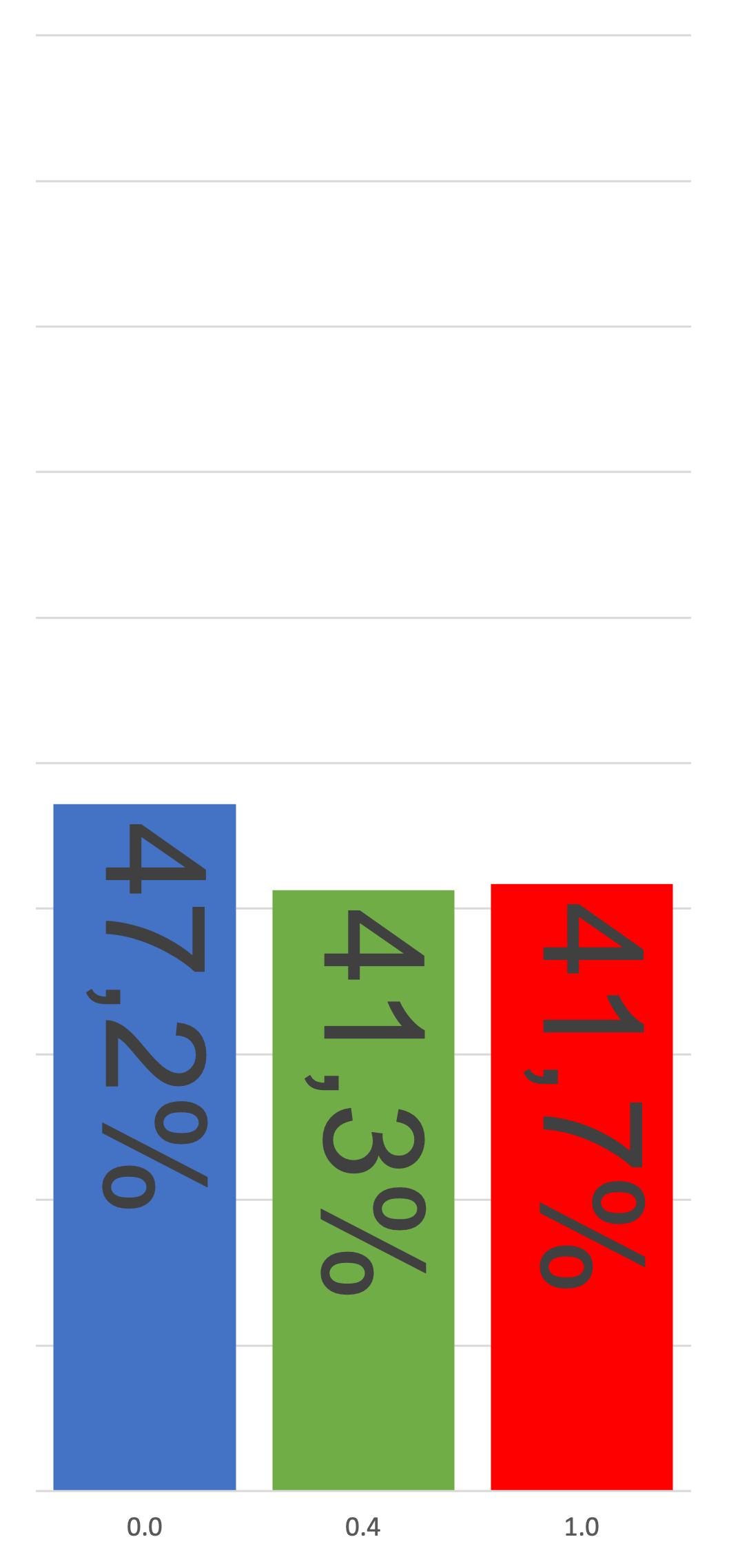} \end{minipage}
			& \begin{minipage}{.14\textwidth} \includegraphics[width=15mm]{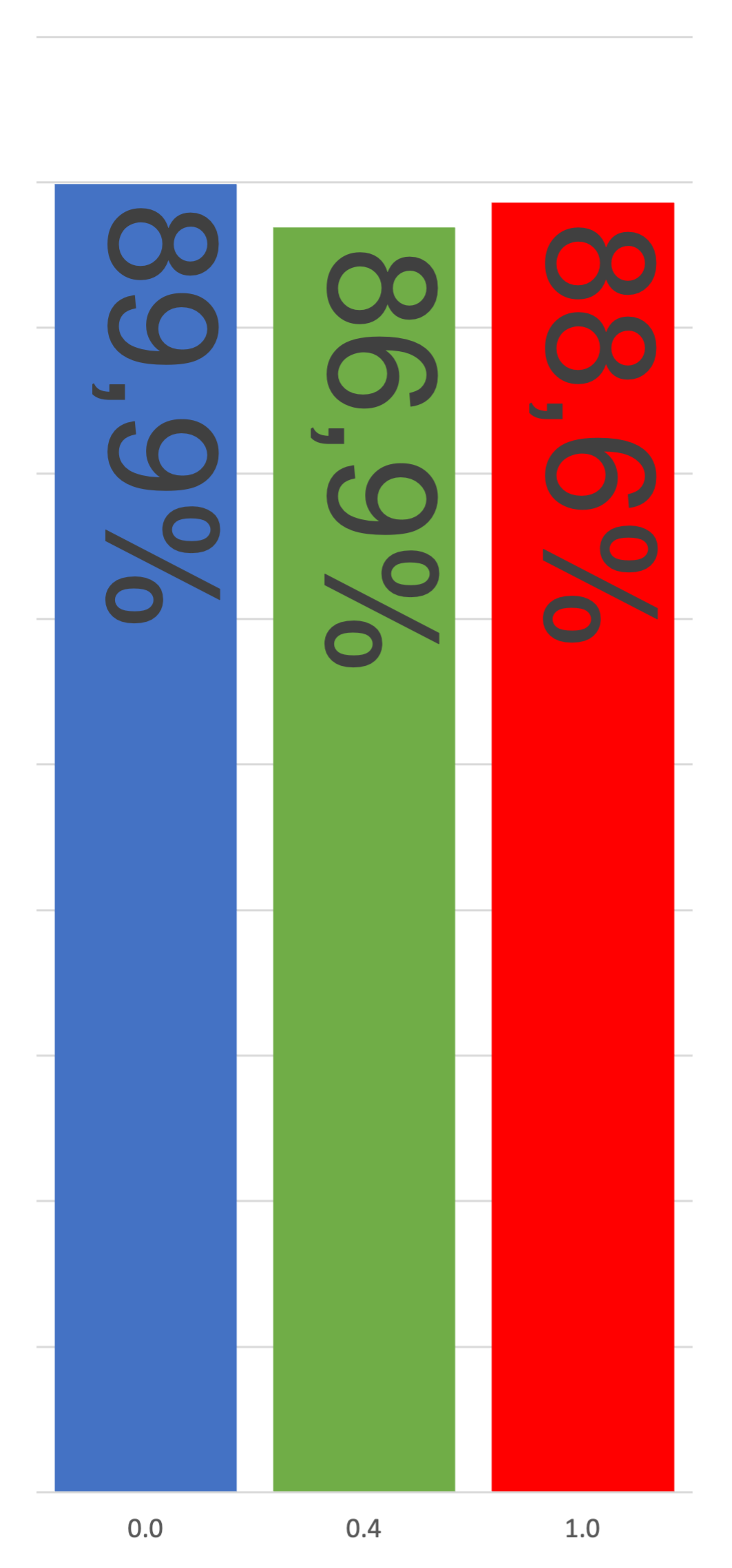} \end{minipage}\\
                \hline    
		\end{tabular}
  }
	\end{threeparttable}
            \caption{Performance measures of the model using all five prompt patterns in requirements traceability} 
		\label{tab:reqtrace}
		\vspace{-.3cm}
\end{table*} 

\begin{table}
\centering
\begin{tabular}{|c|c|c|c|c|}
\hline
Binary Requirements Classification & P-STDEV & R-STDEV & F-STDEV & A-STDEV \\ \hline
Cognitive Verifier & 1,9\% & 2,0\% & 1,7\% & 1,7\% \\
\hline
Context Manager & 3,7\% & 13,5\% & 7,7\% & 2,0\% \\
\hline
Persona & 5,4\% & 0,1\% & 2,8\% & 3,3\% \\
\hline
Question Refinement & 2,0\% & 2,5\% & 0,5\% & 0,8\% \\
\hline
Template & 5,3\% & 0,8\% & 3,5\% & 4,8\% \\
\hline
\end{tabular}

\caption{Standard deviation of performance measures for the five prompt patterns in binary requirements classification task.}
    \label{tab:reqclassstdev}
\end{table}

\begin{table}
\centering
\begin{tabular}{|c|c|c|c|c|}
\hline
Tracing Dependant Requirements & P-STDEV & R-STDEV & F-STDEV & A-STDEV \\ \hline
Cognitive Verifier & 3,7\% & 1,0\% & 3,0\% & 1,1\% \\
\hline
Context Manager & 2,5\% & 4,5\% & 2,1\% & 1,4\% \\
\hline
Persona & 3,8\% & 2,7\% & 3,9\% & 1,4\% \\
\hline
Question Refinement & 1,2\% & 1,7\% & 1,1\% & 0,3\% \\
\hline
Template & 3,9\% & 3,3\% & 3,3\% & 1,5\% \\
\hline
\end{tabular}
\caption{Standard deviation of performance measures for the five prompt patterns in requirements traceability task.}
    \label{tab:reqtracestdev}
\end{table}

Table~\ref{tab:reqclass} and Table~\ref{tab:reqtrace} present the performance measures of the model using all five prompt patterns in the binary requirements classification task and requirements traceability task respectively. The red column represents the value of the metric for the temperature setting of 1.0, the green column represents the value of the metric for the temperature setting of 0.4 and the blue column represents the value of the metric for the temperature setting of 0.0. Table~\ref{tab:reqclassstdev} and Table~\ref{tab:reqtracestdev} present the standard deviations of precision, recall, F-Score and accuracy measures across the three temperature settings, denoted by P-STDEV, R-STDEV, F-STDEV and A-STDEV respectively.

The precision, recall, and F1 scores, which serve as crucial indicators of the model's performance, consistently exhibit higher values in the binary requirements classification task as compared to the tracing of dependent-requirements task. Observations from the results presented above underscore the model's ability to discern and classify binary requirements effectively, demonstrating a higher precision in isolating relevant instances, and a better recall in identifying all pertinent cases. A major observation deviation from this is the higher accuracy scores the model achieved for all patterns for the requirements traceability task in comparison to the binary requirements classification task. This rather significant deviation might be the result of the model being more adept at accurately predicting false negatives than identifying true positives, true negatives and false positives. This leads to a scenario where not wrongly predicting non-existing trace links between two requirements within the SRS documents results in higher accuracy scores. This is why we focus our analysis and base our observations more on precision, recall and F-score measures than accuracy. 

\subsubsection{Cognitive Verifier Pattern:}

From Table~\ref{tab:reqclass}, we can see that the recall is higher than precision when implementing this pattern for binary requirements classification. This holds true for all three temperature settings. This means, the model is observed to be more adept at categorizing NFRs accurately as NFRs (predicting true positives) than making sure the categorized NFRs are indeed NFRs (predicting true positives as well as false negatives) using the Cognitive Verifier pattern.

The variability in standard deviation scores, as seen in Table~\ref{tab:reqclassstdev}, suggests that the effectiveness of these patterns is sensitive to the temperature setting. A higher standard deviation score indicates lower dependability in the model's classification results at varying temperature settings. The standard deviation scores of precision, recall, F-score and accuracy are no more than two percentage points for this pattern. This means that the effect of temperature on the model's performance is low when using this pattern, indicating a higher dependability in binary requirements classification.

When it comes to the requirements traceability task, this pattern yielded higher precision, recall, F-score and accuracy at higher temperature settings, as seen in Table~\ref{tab:reqtrace}. The standard deviation of recall and accuracy is observed to be lower compared to the standard deviation of precision and F-score as observed in Table~\ref{tab:reqtracestdev}. This indicates this pattern is not as good at performing requirements traceability as it is at binary requirements classification.

\subsubsection{Context Manager Pattern:}

From Table~\ref{tab:reqclass}, we observe that precision increases by 6\% between low and high-temperature settings while recall sees a significant drop from 73\% to 49\%. The F-Score and Accuracy also drop with an increase in temperature but not as significantly as recall. The standard deviation scores of precision, recall, F-score and accuracy also vary significantly as in from Table~\ref{tab:reqclassstdev} and Table~\ref{tab:reqtracestdev}. These inconsistent results and high variability across standard deviation measures indicate that this pattern is not a very dependable pattern to use when performing binary requirements classification.

When looking at the requirements traceability task, this pattern yielded the lowest precision, recall and F-score values among all the patterns. Only the accuracy scores are somewhat closer to the accuracy values of other patterns. Therefore, this pattern might not be a suitable choice for performing requirements traceability either.

\subsubsection{Persona Pattern:}

The performance of the model when using the Persona pattern is better than the Context Manager pattern but not as good as the Cognitive Verifier pattern. It has a higher precision value compared to the Cognitive Verifier but the standard deviation of precision values is almost three times that of the Cognitive Verifier pattern. This suggests that when the temperature is not known or adjustable, the Persona pattern might not be the most dependable if precision is more important. A major observation is the almost negligible effect the temperature setting has on the recall, standing at 0.1\%. This means the Persona pattern has the ability to make the model predict true positives with the same level of accuracy with little to no impact from varying the model's temperature.

The Persona pattern yielded the highest recall scores compared to the other patterns when it comes to requirements traceability task. The standard deviation of precision, recall and F-score is between 3\%-4\% which is the second highest among the chosen patterns. The Persona pattern does not seem to be a great choice when performing requirements tracing unless the focus is solely on achieving a high recall. Even then, we recommend a lower model temperature setting to achieve the best results. 

\subsubsection{Question Refinement Pattern:}

The Question Refinement Pattern yielded the highest average precision scores in comparison to other patterns. The recall, F-score, and accuracy are slightly lower than the Cognitive Verifier pattern but also slightly higher than the other patterns. The standard deviation of precision and recall are around 2\% and 2.5\% but for F-score and accuracy, they are under 1.0\%. This means this pattern can be considered a dependable pattern to use when performing binary requirements classification with a statistically insignificant effect of varying the model's temperature.

Even in the requirements traceability task, this pattern yielded the highest average precision, F-score and accuracy values in comparison to other patterns. It also achieved the lowest standard deviation scores for all the performance metrics, making it the top choice for implementation for requirements traceability task.

\subsubsection{Template Pattern:}

Finally, coming to the Template pattern, it has the highest observed recall measures across all three temperature settings among all patterns tested. The standard deviation of recall is also very low, indicating the model can predict true positives with the same level of accuracy with little to no impact from varying the model's temperature. However, the precision and accuracy are below 70\% at default and higher temperature settings. It has a significant amount of standard deviation for precision, F-score and accuracy across temperatures. This makes us question the pattern's ability to make the model yield consistent results when performing binary requirements classification.

The template pattern does not show any noteworthy improvement in results when it comes to the requirements traceability task. The standard deviation scores of precision, recall and F-score are all above 3\% with none of the performance metrics achieving higher scores in comparison to other patterns.

\subsection{Recommendations} \label{subsec:rq2}

Based on our analysis of the results presented in Table~\ref{tab:reqclass}, we can say that the \textbf{Cognitive Verifier} pattern and the \textbf{Question Refinement} pattern are better suited for the binary requirements classification task at any temperature setting compared to the \textbf{Context Manager}, the \textbf{Template} and the \textbf{Persona} patterns.

Similarly, we can say that the \textbf{Question Refinement} pattern is the most consistent and reliable pattern amongst the five for tracing dependent requirements followed by the \textbf{Cognitive Verifier} and the \textbf{Persona} pattern. The \textbf{Context Manager} pattern and the \textbf{Template} pattern are the least reliable pattern for this task.

\begin{table}[]
    \centering
    \begin{tabular}{|l|l|l|l|}
    \hline
     Rank & Binary Classification  & Tracing & Overall\\
     \hline
       1st  & Question Refinement & Question Refinement & Question Refinement \\
       2nd & Cognitive Verifier & Cognitive Verifier & Cognitive Verifier\\
       3rd & Persona & Persona & Persona\\
       4th & Template & Template & Template\\
       5th & Context Manager & Context Manager & Context Manager\\
       \hline
    \end{tabular}
    \caption{Rank Based Prompt Pattern Recommendation for Overall and Individual Tasks}
    \label{tab:prompt_rec}
\end{table}

Overall, the \textbf{Question Refinement} pattern, shows consistent results across both the classification and requirements traceability tasks. The \textbf{Cognitive verifier} pattern and the \textbf{Persona pattern} obtained higher performance scores in binary requirements classification, although their performance in tracing dependent requirements was reduced. The \textbf{Context Manager} pattern was found to have a greater degree of variability in its STDEV measures for both tasks. Our results indicate this pattern may not be the best-suited pattern for performing the selected requirements engineering tasks.

\subsection{Evaluation framework} \label{subsec:rq3}

In order to evaluate the effectiveness of prompt patterns for any RE task, we propose adopting a framework similar to the methodology used in this paper.

\begin{itemize}
    \item \textbf{Step-1:} Curate a dataset for the task in question, comprising two distinct versions. The first version should contain the ground truth annotations, while the second version should be cleaned to remove any identifiers that helped in establishing the ground truth. This version will serve as input data provided to the GenAI model through a program script.
    \item \textbf{Step-2:} Create a program script\footnote{\url{https://github.com/beatrizcabdan/GenAI4REtasks}} that mimics the RE task's underlying logic. The script should be designed to leverage the capabilities of the GenAI model via an API call function. Clearly specify the desired output format within the prompt embedded in the code script. Ensure that the script is capable of taking the second version of the dataset as input and generating results using the GenAI model.
    \item \textbf{Step-3:} Execute the code script created in the previous step to generate results. The script should make API calls to the GenAI model, using the specified prompt pattern.
    \item \textbf{Step-4:} Conduct a comparative analysis of the obtained results against the ground truth annotations (from the first version of the dataset). This analysis will provide insights into how well the GenAI model performed in relation to the ground truth.
    \item \textbf{Step-5:} Use the comparative assessment results to evaluate the effectiveness of the prompt pattern(s) in the context of the specific RE task.
\end{itemize}

\begin{figure}[ht!]
   \centering
   \includegraphics[width=\linewidth]{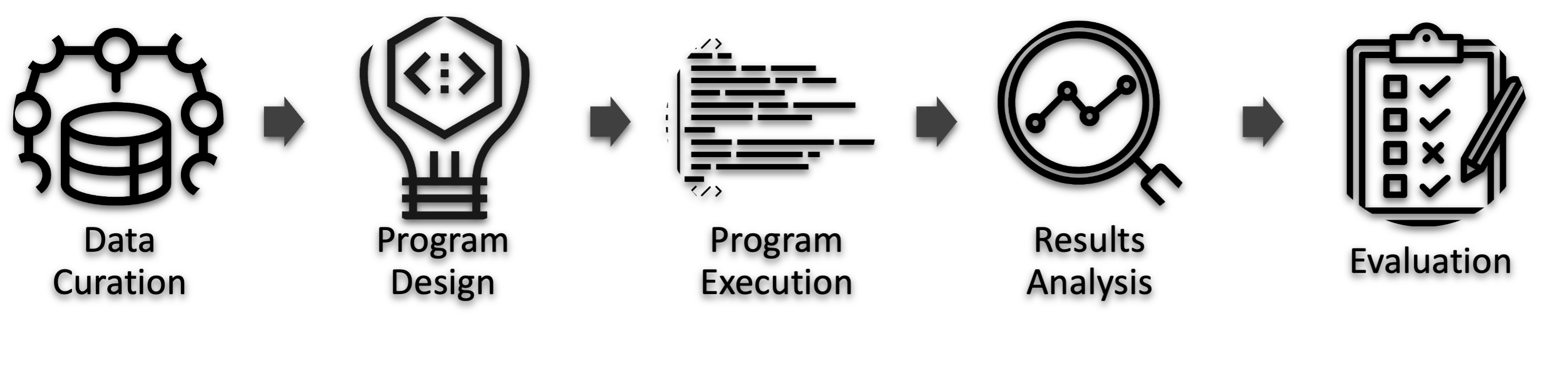}
   \caption{Prompt Pattern Effectiveness Evaluation Framework}
   \label{fig:fig-1}
\end{figure}

By analysing the requirements and objectives of the task, one can determine whether the nature of the task leans more towards binary classification or requirements traceability. Based on this, our recommendations for which prompt patterns to apply may hold. However, the results of this study may not consistently apply to other situations. When confronted with a novel RE task that cannot be framed as a requirements IR task, it is advisable to experiment with the patterns and evaluate the effectiveness of patterns in the context of the specific task and dataset as suggested in our evaluation framework.

\section{Discussion}

The results of our research study shed light on the effectiveness of different prompt patterns in the context of RE tasks, specifically focusing on binary requirements classification and requirements traceability. In the following, we will delve into the implications of these findings and their practical applications in real-world practice.

Our study revealed that the \textbf{Cognitive Verifier} and \textbf{Question Refinement} patterns achieved the best results in binary requirements classification. These patterns provide a reliable and consistent approach to achieving accurate and reliable classifications. In real-world RE practice, RE practitioners can consider adopting either the \textbf{Cognitive Verifier} or \textbf{Question Refinement} pattern for tasks that involve binary classification. For instance, when evaluating software requirements for compliance with specified standards, these patterns could be used to streamline the classification process, reducing manual effort and potential errors. The \textbf{Persona} pattern seems to exhibit better results at lower temperature settings compared to higher temperature settings, indicating it is better suited to classification tasks where less creative and more definitive responses are required.

In the case of requirements traceability, our findings indicate that the \textbf{Question Refinement} pattern outperforms others. This suggests that when the RE task involves establishing relationships and dependencies between various requirements, using the \textbf{Question Refinement} pattern is the most effective option. For RE teams tasked with tracing dependencies among requirements, the \textbf{Question Refinement} pattern has significantly better performance compared to other patterns. This is particularly valuable in complex projects where understanding how changes in one requirement may impact others is critical.

Since the performance measures were calculated automatically using the dependencies provided in the datasets, we did not manually investigate the requirements that were misidentified as dependent (or independent). Further work, particularly work focusing specifically on traceability, should look closer to investigate the reasons behind the misclassification. The sensitivity of the prompt used to generate some artefacts using LLMs or any other GenAI model opens up possibilities where minimal changes to the prompt can result in significant differences in the quality of the output as well as the performance of the model. This experimental study was limited to examining the effect of pattern-level variations on the prompts used and did not look into the specific wording of the prompts, in order to keep the experimental results tractable. It is equally important to remember that using LLMs for RE tasks should be limited to assisting relevant RE stakeholders with appropriate human oversight mechanisms in place instead of automating these tasks. Therefore, the usefulness of LLMs, and subsequently, prompt patterns that are used to craft the prompts to interact with the LLMs, comes with limitations and needs more dedicated research results to discuss it in depth.

While our study provides insights into the effectiveness of specific patterns, it is essential to acknowledge the unique nature of various RE tasks. Not all tasks can be framed as requirements IR tasks. Therefore, organizations should consider an analysis of their specific RE requirements and objectives. When confronted with novel RE tasks, teams can follow a structured approach similar to our proposed framework presented in Figure~\ref{fig:fig-1} to identify the most suitable prompt pattern. By breaking down the workflow and considering the task's nature, they can adapt and experiment with different patterns to optimize results. This iterative process allows for continuous improvement in the choice of patterns for specific tasks.

\section{Conclusion}

In conclusion, the insights garnered from this study offer guidance for practitioners seeking to leverage prompt patterns for using GenAI in RE tasks. Our research offers recommendations on the selection and adoption of prompt patterns for real-world RE tasks. We suggest that the Question Refinement pattern might serve as a suitable compromise for both tasks. Moreover, the paper presents an evaluation framework based on the methodology used in our study on how one might evaluate and decide which prompt pattern could be the most effective for a new RE task, one that considers the trade-offs between precision, recall, and accuracy. Practitioners/other researchers can use this framework as a guideline for assessing the suitability of prompt patterns for their unique RE tasks. By understanding the strengths and limitations of different patterns and employing a structured evaluation framework, organizations can enhance the efficiency and accuracy of their RE processes, ultimately leading to improved software development outcomes and project success.

The insights presented can lay the foundation for several avenues of future research, aiming to deepen our understanding of prompt patterns and further enhance the performance of GenAI in RE. Future investigations could delve into a more exhaustive exploration of prompt patterns, potentially identifying novel approaches. An ensemble approach, combining the merits of different patterns, may mitigate the limitations associated with individual patterns and contribute to a more robust and adaptable classification framework. Future research could focus on optimizing the balance, exploring strategies that prioritize comprehensive recall without compromising precision or accuracy which could lead to more context-aware and adaptable models and synthesizing new prompt patterns and evaluation framework. Another avenue of further research is to establish the boundary of LLMs' application in RE activities with appropriate human oversight mechanisms in place to ensure the ethical and responsible application of these technologies.

\section{Acknowledgement}
This work was supported by the Vinnova project ASPECT [2021-04347].

\bibliographystyle{splncs04}
\bibliography{main}

\end{document}